\documentclass[aps,prb,twocolumn,10pt,floatfix,flushbottom]{revtex4-2}
\usepackage{adjustbox}
\usepackage[utf8]{inputenc}
\usepackage[colorlinks=true,linkcolor=magenta,urlcolor=blue,citecolor=blue]{hyperref}
\usepackage{mathtools,amsmath,amssymb,amsfonts,graphicx,tabularx,dcolumn,bm,xcolor,times,float,algorithm,algorithmic,tikz,pgf}
\usepackage{dsfont}
\usepackage{capt-of}
\usepackage{multirow}
\usepackage{CJKutf8}
\usepackage{subfig} 
\usepackage[justification=justifying]{ragged2e}

\makeatletter
\newcommand{\II}{\mathbb{I}\,}
\newcommand{\XX}{\mathbb{X}}
\renewcommand{\Re}{\operatorname{Re}}
\makeatother

\DeclareMathOperator{\tr}{tr}
\usetikzlibrary{arrows.meta, decorations.markings,calc}
\usetikzlibrary{positioning}
\tikzstyle{vecArrow} = [thick, decoration={markings,mark=at position
   1 with {\arrow[semithick]{open triangle 60}}},
   double distance=1.4pt, shorten >= 5.5pt,
   preaction = {decorate},
   postaction = {draw,line width=1.4pt, white,shorten >= 4.5pt}]
\tikzstyle{innerWhite} = [semithick, white,line width=1.4pt, shorten >= 4.5pt]
\tikzset{middlearrow/.style={
        decoration={markings,
            mark= at position 0.5 with {\arrow{#1}} ,
        },
        postaction={decorate}
    }
}

\begin{document}
\begin{CJK*}{UTF8}{gbsn} 
\title{Operator Spreading in Random Unitary Circuits with Unitary-invariant Gate Distributions}
\author{Zhiyang Tan (谭志阳)}
\author{Piet W.\ Brouwer}%
\affiliation{Dahlem Center for Complex Quantum Systems and Physics Department, Freie Universit\"at Berlin, Arnimallee 14, 14195 Berlin, Germany}
\date{\today}
\begin{abstract}
Random unitary circuits have become a model system to investigate information scrambling in quantum systems.
In the literature, mostly random circuits with Haar-distributed gate operations have been considered.
In this work, we investigate operator spreading in random unitary circuits in which the elementary gate operations are drawn from general unitary-invariant ensembles, which include the well-studied Haar-distributed random unitary circuits as a special case. Similar to the Haar-distributed case, the long-time behavior of operator spreading with the more general unitary-invariant gate distribution is governed by drift-diffusion equations characterized by the butterfly velocity $v_{\rm B}$ and a diffusion constant $\mathcal{D}$. Differences with the Haar-random case are (i) that it takes a finite time $\tau_{\rm b}$ until ensemble-averaged Pauli-string weights take a ``binary'' form, in which they depend only on whether Pauli operators inside the support of the Pauli strong are equal to the identity matrix, and (ii) that the operator spreading is characterized by a finite ``domain-wall width'' $n_{\rm DW}$ separating regions with a random-matrix-like Pauli-string distribution.
To illustrate these findings, we perform explicit calculations for random unitary circuits distributed according to the Poisson kernel, which interpolates between the trivial and Haar-distributed circuits.
\end{abstract}
\maketitle
\end{CJK*}
\section{Introduction}
\begin{figure}
\centering
\includegraphics[scale=0.48,trim={4cm 2cm 2cm 2cm},clip]{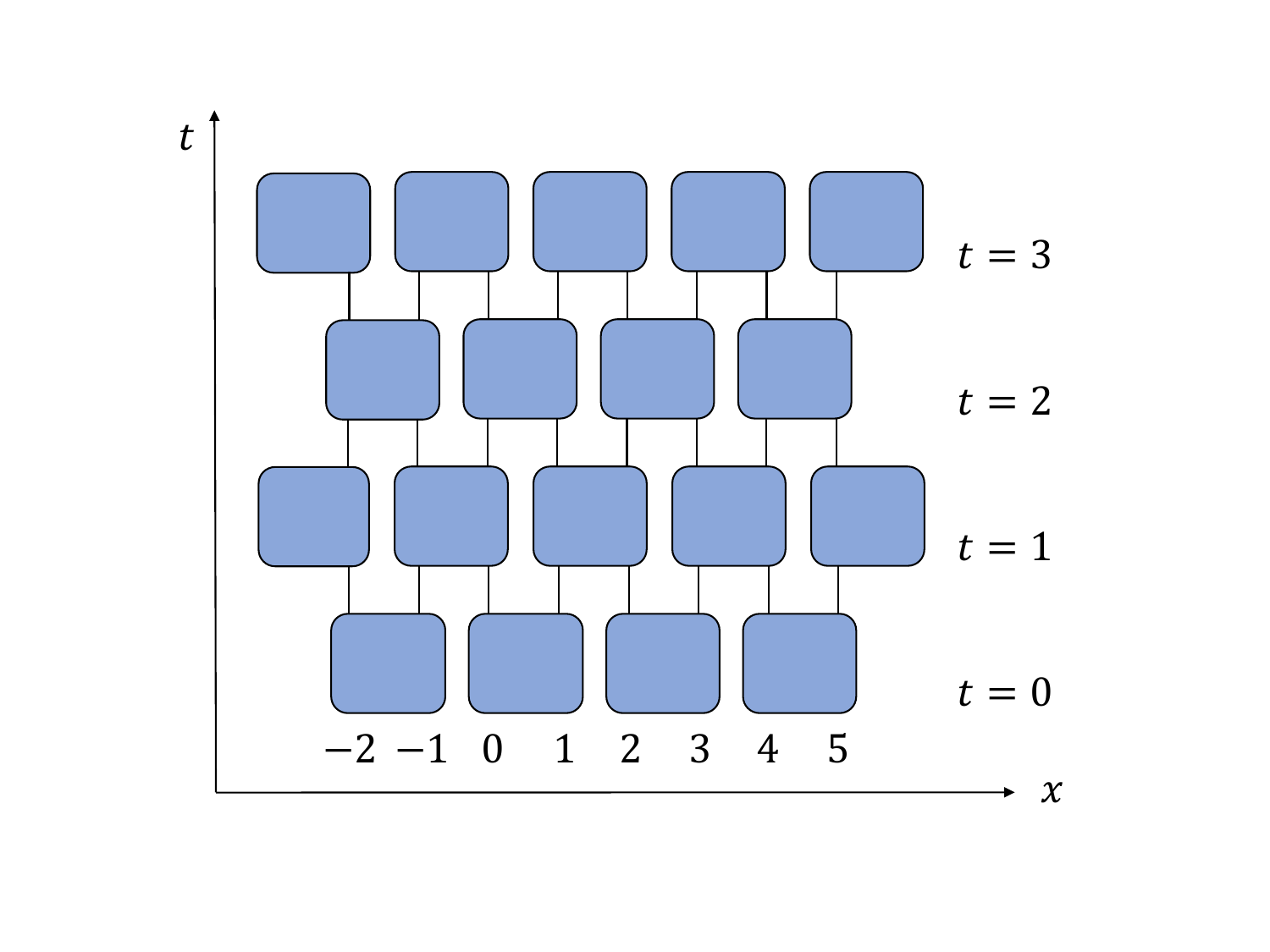}
\caption{\justifying \small
Sketch of a random unitary circuit. It has a brickwork structure and is composed of gates uncorrelated both in space 
and time. }
\label{Fig:circuit}
\end{figure}

In recent years, there has been a growing interest in the study of random unitary circuits \cite{brown2013scrambling, Nahum2017quantum, chan2018solution, nahum_operator_2018, von_keyserlingk_operator_2018, rakovszky_diffusive_2018, khemani_operator_2018, Fisher2023random, skinner2023lecture}. Random unitary circuits provide a minimal model in which information spreading, quantum chaos, and thermalization in quantum-mechanical systems can be studied. A typical random circuit consists of a sequence of gate operations acting on pairs of $q$-dimensional quantum degrees of freedom (``qudits''), whereby the gate operations are drawn from a statistical distribution \cite{Nahum2017quantum, nahum_operator_2018, von_keyserlingk_operator_2018}. The gate operations follow a ``brickwork'' structure, in which the pattern of gates that are acted upon by the gate operation alternates in space and time, see Fig. \ref{Fig:circuit}.

Quantum information scrambling \cite{xu_locality_2019, styliaris_information_2021} refers to the phenomenon that a local operator $O$, as it evolves in time, acquires non-local correlations \cite{xu_scrambling_2022}. These non-local correlations can be described by expressing the Heisenberg-picture operator $O(t)$ (with $O(0) = O$) in terms of a set of Pauli-string basis operators $O_p$, which are products of generalized Pauli operators $\sigma_{p_x}$ associated with each qudit, and the corresponding amplitudes $\gamma_p(t)$,
\begin{equation}
  O(t) = \sum_{p} \gamma_p(t) O_p. \label{eq:Ot}
\end{equation}
Quantum information scrambling then arises from two primary effects: operator spreading, which is the increase of the support of the Pauli strings $p$ appearing in the sum (\ref{eq:Ot}) with time $t$, and entanglement spreading, which is the increase of the number of terms in the sum (\ref{eq:Ot}) with time \cite{xu_scrambling_2022,mi_information_2021}. (The Pauli string index $p$ is a composite index $p = [\ldots,p_{x-1},p_{x},p_{x+1},\ldots]$. Its support is the set of qudit positions $x$ for which $\sigma_{p_x}$ is different from the identity operator.)

The random unitary circuit with maximal randomness (given the brickwork structure) has its two-qudit gate operations ${\cal U}$ drawn from a Haar distribution \cite{nahum_operator_2018, von_keyserlingk_operator_2018, rakovszky_diffusive_2018, khemani_operator_2018}. The Haar distribution is uniform across the unitary group,
\begin{equation}
  \label{eq:PHaar}
  P_{\rm Haar}({\cal U}) = \mbox{const}.
\end{equation}
For such a Haar-distributed random unitary circuit, Nahum {\em et al.} \cite{nahum_operator_2018} characterized operator spreading by the ensemble average $\rho_p(t) \equiv \langle |\gamma_p(t)|^2 \rangle$, for which they found that the endpoints $x_p$ of the support of typical Pauli strings $p$ contributing to $\rho_p(t)$ satisfied a drift-diffusion equation with drift velocity 
\begin{equation}
  v_{\rm B} = \frac{q^2-1}{q^2+1}
  \label{eq:vBHaar}
\end{equation}
and diffusion constant
\begin{equation}
  {\cal D} = \frac{4 q^2}{(q^2+1)^2}.
  \label{eq:DHaar}
\end{equation}
The drift velocity is commonly referred to as the {\em butterfly velocity}. It approaches the light-cone limit $v_{\rm B} = 1$ in the limit of large qudit dimension $q$.

In this article, we consider operator spreading in a class of random unitary circuits with non-Haar distributed two-qudit gates, while still requiring that the probability distribution for a two-qudit gate operation ${\cal U}$ is invariant under unitary transformations, {\em i.e.},
\begin{equation}
  P({\cal U}) = P({\cal V} {\cal U} {\cal V}^{\dagger})
  \label{eq:unitary_invariance}
\end{equation}
for arbitrary unitary matrices ${\cal V}$.
A special case of such a unitary-invariant distribution is the Poisson kernel \cite{KRIEGER1967375,forrester2010,brouwer_generalized_1995,mello1985a}
\begin{equation}
  P_{\rm Poisson}({\mathcal U};\alpha) \propto
  |\det(\mathds{1} - \alpha {\mathcal U})|^{-2 q^2},
  \label{eq:Poisson}
\end{equation}
which is the maximum-information-entropy distribution with the constraint that the ensemble average
\begin{equation}
  \langle {\cal U} \rangle = \alpha \mathds {1}.
\end{equation}
The Poisson kernel distribution allows a continuous interpolation between the limit $|\alpha| = 1$ of a circuit with no information spreading and the maximally random Haar limit $\alpha = 0$.

Like for Haar-distributed random unitary circuits \cite{nahum_operator_2018}, operator spreading in random unitary circuits with the invariance property (\ref{eq:unitary_invariance}) can be mapped to a classical stochastic growth problem, which in the long-time limit approaches a drift-diffusion process with butterfly velocity $v_{\rm B}$ and a diffusion constant ${\cal D}$. In addition to $v_{\rm B}$ and ${\cal D}$ being different from those for the Haar limit, the unitary-invariant random circuit differs from the maximally random Haar-distributed random circuit in two more ways: (i) For the Haar-random circuit, the ensemble-averaged Pauli string weights $\rho_p(t)$ depend only on whether or not each generalized Pauli operator $\sigma_{p_x}$ equals the identity operator. For the unitary-invariant circuit studied here, such a ``binary'' form of the Pauli string weights $\rho_p(t)$ sets in only after a time $\tau_{\rm b}$. (ii) For the Haar-random circuit, Pauli operators inside the support of a Pauli string have a random-matrix distribution, in which all Pauli operators have the same probability. For the unitary-invariant circuits, this random-matrix limit sets in only at a finite ``domain wall width'' $n_{\rm DW}$ from the end of the Pauli string. For a random circuit with Poisson-kernel-distributed two-qubit gate operators, we find that both $\tau_{\rm b}$ and $n_{\rm DW}$ are finite, although $\tau_{\rm b}$ diverges in the limit $|\alpha|\to 1$ of a trivial circuit. 

The remainder of this article is organized as follows: In Sec.\ \ref{sec:2} we briefly review the formulation of random unitary circuits with the brickwork structure of Fig.\ \ref{Fig:circuit}, closely following Ref.\ \cite{nahum_operator_2018}. We map to a stochastic growth model and obtain the butterfly velocity $v_{\rm B}$ and diffusion constant ${\cal D}$ for random circuits with unitary-invariant two-qudit-gate distributions in Sec.\ \ref{sec:3}. Results for the Poisson-kernel distribution are presented in Sec.\ \ref{sec:4}. In Sec.\ \ref{sec:5}, we show that the Pauli-string weights $\rho_p$ acquire a binary form after a finite time $\tau_{\rm b}$ and calculate the out-of-time-ordrered correlator (OTOC) for the type of random circuits considered here. We conclude in Sec.\ \ref{sec:concl}. The appendices contain additional details of our calculations.

\section{Random unitary circuit}
\label{sec:2}

\subsection{Pauli string coefficients}

We consider a one-dimensional array of $q$-state qudits of length $L$ and periodic boundary conditions labeled by a site index $x$, shown schematically in Fig.\ \ref{Fig:circuit}. A complete operator basis for the qudit at site $x$ is formed by generalized Pauli matrices $\sigma_{p_x}$, where $p_x = 0, \ldots, q^2-1$ and $\sigma_{0}$ is the identity operator. A basis for operators on the full one-dimensional array is given by so-called ``Pauli strings'' $O_{p}$, tensor products of the generalized Pauli operators associated with each qudit. A Pauli string is labeled by the composite index $p = [p_{0},p_{1},p_{2},\ldots,p_{L-1}]$, which contains one entry $p_x$ for each qudit. The Pauli-string basis operators satisfy the orthonormality property \cite{roberts_chaos_2017}
\begin{equation}
  \frac{1}{q^L} \mbox{tr}\, O_p O_q^{\dagger} = \delta_{p,q}.
\end{equation}
For more details on generalized Pauli operators, we refer to App. \ref{app:1}.

\subsection{Time evolution}

Time evolution takes place via the unitary evolution operator $U(t,t')$,
\textcolor{black}{
\begin{equation}
  O(t) =  U^{\dagger}(t,t') O(t')U(t,t').
  \label{eq:UO}
\end{equation}}
Following Refs.\ \cite{nahum_operator_2018, hunter-jones_operator_2018}, we consider a stochastic stroboscopic evolution, which is a product of two-qudit gate operators,
\begin{equation}
  \label{eq:gates}
  U(t,t-1) =
  \left\{ \begin{array}{ll}
    \otimes_{\textrm{$x$ even}}\, {\cal U}_{x,x+1} & \mbox{for $t$ even}, \\
    \otimes_{\textrm{$x$ even}}\, {\cal U}_{x-1,x} & \mbox{for $t$ odd},
  \end{array} \right.
\end{equation}
where ${\cal U}_{x,y}(t,t-1)$ is a $q^2 \times q^2$ matrix that acts on the qudit degrees of freedom of sites $x$ and $y$ only. This structure is shown schematically in Fig.\ \ref{Fig:circuit}.

In Refs.\ \cite{nahum_operator_2018, hunter-jones_operator_2018}, for each time step, the matrices ${\cal U}_{x,x+1}(t,t-1)$ or ${\cal U}_{x-1,x}(t,t-1)$ are taken independently from the group of $q^2 \times q^2$ unitary matrices according to the uniform distribution $P_{\rm Haar}$, see Eq.\ (\ref{eq:PHaar}). Here, we still take the matrices ${\cal U}_{x,x+1}(t,t-1)$ or ${\cal U}_{x-1,x}(t,t-1)$ from independent distributions, but for the form of the distribution we only assume invariance with respect to unitary basis transformations, see Eq.\ (\ref{eq:unitary_invariance}).
\subsection{Time evolution of Pauli-string coefficients}

The time evolution of the unitary circuit starts from a local operator $O$ at time $t=0$, such that $O$ equals the identity operator $\sigma_0$ for sites sufficiently far away from a reference site $x_0$. We require that $O$ satisfies the normalization condition
\begin{equation}
  \frac{1}{q^L} \mbox{tr}\, O O^{\dagger} = 1.
\end{equation}
A Pauli string operator that contains $\sigma_{p_{x_0}}$ at qudit $x_0$ and $\sigma_0$ at all other qudits is an example of an operator that meets this normalization condition. The Heisenberg-picture operator $O(t)$ can be expanded in the basis of Pauli strings as in Eq.\ (\ref{eq:Ot}),
where
\begin{equation}
  \gamma_p(t) = \frac{1}{q^L} \mbox{tr}\, O(t) O_p^{\dagger}.
\end{equation}
The time evolution of the Pauli-string coefficients $\gamma_p$ follows from that of the operator $O$, see Eq.\ (\ref{eq:UO}). Because $O = O(0)$ is normalized and the unitary time evolution (\ref{eq:UO}) preserves the normalization, $\gamma_p(t)$ satisfies the normalization condition
\begin{equation}
  \sum_{p} |\gamma_{p}(t)|^2 = 1
  \label{eq:gamma_normalization}
\end{equation}
for all $t$.

For a single-time step, we have
\begin{align}
  \gamma_{p}(t)  =&\,
  \sum_{a} W_{ap}(t,t-1) \gamma_a(t-1),
  \label{eq:gammat}
\end{align}
where the summation variable $a$ labels Pauli strings and
\textcolor{black}{
\begin{align}
  W_{ap}(t,t-1) = \frac{1}{q^L} 
  \mbox{tr}\, U^{\dagger}(t,t-1) O_{a} U(t,t-1) O_p^{\dagger}.
  \nonumber
\end{align}}
For an evolution matrix of the form (\ref{eq:gates}), this evaluates to
\begin{equation}
  W_{ap} =
  \left\{ \begin{array}{ll}
    \prod_{\textrm{$x$ even}}
    {\cal W}_{ap;x,x+1}
    & \mbox{$t$ even}, \\
    \prod_{\textrm{$x$ even}}
   {\cal W}_{ap;x-1,x} & \mbox{$t$ odd},
  \end{array} \right.
\end{equation}
where the factors ${\cal W}_{ap;x,y}$ do not depend on the full Pauli strings $a$ and $p$, but only on the labels $a_x$, $a_y$, $p_x$, and $p_y$ that belong to the sites $x$ and $y$,
\textcolor{black}{\begin{align}
  {\cal W}_{ap;x,y} =&\,
  \frac{1}{q^2}
  \mbox{tr}\, {\cal U}^{\dagger}_{x,y}
  (\sigma_{a_x} \otimes \sigma_{a_y})
  {\cal U}_{x,y}
  (\sigma_{p_x}^{\dagger} \otimes \sigma_{p_y}^{\dagger}).
  \label{eq:WU}
\end{align}}
The coefficients ${\cal W}_{ap;x,y}$ satisfy the unitarity relations
\begin{align}
  \sum_{p_x,p_y} {\cal W}_{ap;x,y} {\cal W}^*_{bp;x,y}
  =&\, \delta_{a_x,b_x} \delta_{a_y,b_y},\nonumber \\
  \sum_{a_x,a_y} {\cal W}_{ap;x,y} {\cal W}^*_{aq;x,y}
  =&\, \delta_{p_x,q_y} \delta_{q_x,q_y},
  \label{eq:unitarity}
\end{align}
which ensures the normalization of the Pauli string coefficients, see Eq.\ (\ref{eq:gamma_normalization}).
They also have the property that
\begin{equation}
  {\cal W}_{ap;x,y} = \delta_{a_x,0} \delta_{a_y,0}\ \ \mbox{if $p_x = p_y = 0$}
\end{equation}
and, similarly,
\begin{equation}
  {\cal W}_{ap;x,y} = \delta_{p_x,0} \delta_{p_y,0}\ \ \mbox{if $a_x = a_y = 0$},
  \label{eq:special}
\end{equation}
which guarantees that a trivial stretch of a Pauli string remains trivial under time evolution and vice versa.

\section{Mapping to stochastic growth model}
\label{sec:3}
\subsection{Stochastic growth model}

The Pauli-string coefficient $\gamma_p(t)$ contains the full evolution information of the operator $O$ under unitary evolution. Following Ref.\ \cite{nahum_operator_2018}, we here focus on the average of the modulus square
\begin{equation}
  \rho_p(t) = \langle |\gamma_p(t)|^2 \rangle,
\end{equation}
which may be interpreted as a (classical) probability distribution on the set of Pauli strings $p$. With the help of Eq.\ (\ref{eq:gammat}), $\rho_p(t)$ may be expressed in terms of the correlation function $\langle \gamma_{a}(t-1) \gamma_{b}(t-1)^* \rangle$ and the covariance $\langle W_{ap} W_{bp}^* \rangle$,
\begin{equation}
  \rho_p(t) =
  \sum_{a,b} \langle W_{ap}W^{*}_{bp}\rangle \langle\gamma_{a}(t-1)\gamma^*_b(t-1)\rangle.
  \label{eq:gammasqavg0}
\end{equation}
In App.\ \ref{app:2} we show that 
\begin{equation}
  \langle W_{ap} W_{bp}^* \rangle = 0\ \mbox{if $a \neq b$} \label{eq:WW}
\end{equation}
if the two-qudit operators ${\cal U}_{x,y}$ are drawn from a probability distribution with the invariance property (\ref{eq:unitary_invariance}). Hence, Eq.\ (\ref{eq:gammasqavg0}) simplifies to a closed Markovian evolution equation for the averaged Pauli-string weights $\rho_p(t)$,
\begin{align}
  \rho_p(t)
  =&\,
  \sum_{a}\
  \rho_a(t-1)
  \langle |W_{ap}|^2 \rangle.
  \label{eq:gammasqavg}
\end{align}
Equation (\ref{eq:gammasqavg}) describes the time evolution of $\rho_p(t)$ as a Markovian stochastic process with transition probabilities $\langle |W_{ap}|^2 \rangle$.

The transition probabilities $\langle |W_{ap}|^2 \rangle$ in this stochastic growth model factorize in pair contributions from the two-qudit gates,
\begin{align}
  \label{eq:Wevenodd}
  \langle |W_{ap}|^2 \rangle
  =&\,
  \prod_{x\, {\rm even}}
  \left\{ \begin{array}{ll}
    \langle |{\cal W}_{ap;x,x+1}|^2 \rangle & \mbox{for $t$ even}, \\
    \langle |{\cal W}_{ap;x-1,x}|^2 \rangle & \mbox{for $t$ odd}.
  \end{array} \right.
\end{align}
Because of the unitary invariance (\ref{eq:unitary_invariance}) of the distribution $P({\cal U}_{x,y})$ of the two-qudit gate operators, the pair transition probabilities depend on low-order moments of the distribution only. The relevant moments are 
\begin{align}
\begin{split}
  {\cal R}_{1;1} =&\, \langle \mbox{tr}\, {\cal U}_{x,x+1}\, \mbox{tr}\, {\cal U}_{x,x+1}^{\dagger} \rangle,  \\
  {\cal R}_{2;2} =&\, \langle \mbox{tr}\, {\cal U}_{x,x+1}^2\, \mbox{tr}\, {\cal U}_{x,x+1}^{\dagger 2} \rangle,   \\
  {\cal R}_{1,1;1,1} =&\, \langle (\mbox{tr}\, {\cal U}_{x,x+1})^2\, (\mbox{tr}\, {\cal U}_{x,x_1}^{\dagger})^2 \rangle,  \\
  {\cal R}_{1,1;2} =&\, \langle (\mbox{tr}\, {\cal U}_{x,x+1})^2)\, \mbox{tr}\, {\cal U}_{x,x+1}^{\dagger 2} \rangle. \label{eq:R112}
  \end{split}
\end{align}
The moments ${\cal R}_{1,1}$, ${\cal R}_{2,2}$, and ${\cal R}_{1,1,1,1}$ are real; ${\cal R}_{1,1;2}$ is complex. In App.\ \ref{app:2} we show that the pair transition probability $\langle |{\cal W}_{ap;x,y}|^2 \rangle$ is
\begin{align}
  \label{eq:Wgeneral}
  \langle |{\cal W}_{ap;x,y}|^2\rangle =&\,
  \delta_{a,0} \delta_{p,0} 
  +  \frac{(1 - \delta_{a,0})(1 - \delta_{p,0})}{q^4-1}
 \\  &\, ~~ \mbox{} \times
 \left(1 - \frac{\mbox{Re}\, A_1 + A_2 + A_3}{q^4-1}
   \right. \nonumber \\ &\, \left. \vphantom{\frac{1}{2}} \mbox{}
 + \mbox{Re}\, A_1 \varphi(a,p)
   + A_2 \delta_{a,-p} + A_3 \delta_{a,p} \right) ,\nonumber 
\end{align}
where $A_1$ is a complex coefficient and $A_2$ and $A_3$ are real coefficients that are linearly independent combinations of the moment functions (for details, we refer to APP. \ref{app:2}).

\begin{figure}[tb]
\centering
\hspace{-0.6 cm}
\includegraphics[scale=0.5
]{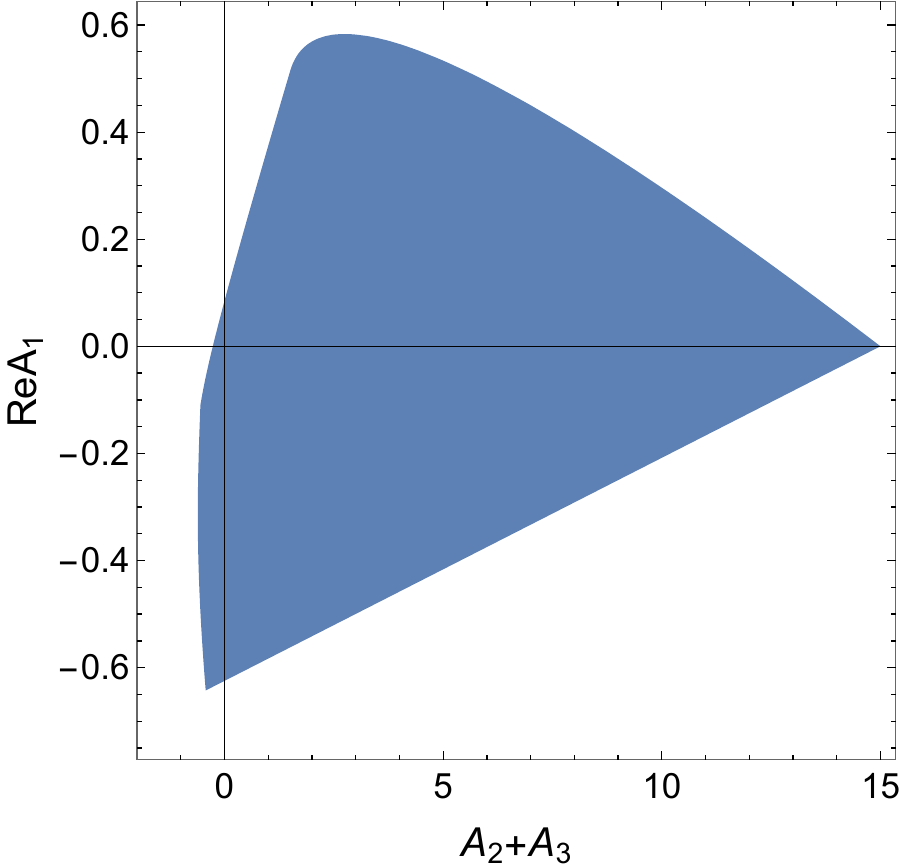} 
\caption{\justifying \small
Parameter space showing the allowed values of \(A_2 + A_3\) and \(\mathrm{Re} A_1\) for $q=2$. The parameter space is the convex enclosure of all points in the $A_2 + A_3, \mathrm{Re}\, A_1$-plane corresponding to moments of unitary matrices ${\cal U}_{x,y} = {\cal V} {\cal U}_0 {\cal V}^{\dagger}$, with ${\cal V}$ Haar-distributed and ${\cal U}_0$ fixed. The parameter space in the figure is obtained by numerically sampling ${\cal U}_0$ uniformly.}
\label{fig:parameter space}
\end{figure}

The function $\varphi(a_x,a_y,p_x,p_y)$ depends on the choice of the basis of generalized Pauli operators. An explicit expression is given in App.\ \ref{app:1} for the choice of the generalized Pauli basis made there. 
The limit of Haar-random two-qudit operators ${\cal U}_{x,y}$ corresponds to $A_1 = A_2 = A_3 = 0$, whereas $A_1 = A_2 = 0$ and $A_3 = q^4-1$ for the trivial limit ${\cal U}_{x,y} = \mathds 1$. 
In Subsec.\ \ref{subsec:3D} it will be shown that only \(\mathrm{Re}\, A_1\) and \(A_2 + A_3\) govern operator spreading at sufficiently large times and, hence, enter into the calculation of the butterfly velocity $v_{\rm B}$ and the diffusion constant ${\cal D}$. In Fig.\ \ref{fig:parameter space} we show the allowed parameter space for these two parameters of the stochastic growth model.
\subsection{Butterfly velocity}

Using the properties (\ref{eq:unitarity})--(\ref{eq:special}), one verifies that the evolution equation (\ref{eq:gammasqavg}) admits the maximally random steady-state solution
\begin{equation}
  \label{eq:rhomax}
  \rho^{\infty}_p(t) = q^{-2 L}
\end{equation}
as well as the trivial solution
\begin{equation}
  \label{eq:rhotrivial}
  \rho^0_p(t) =
  \prod_{x} \delta_{p_x,0}.
\end{equation}
Except in the case of a trivial circuit, these are the only two stationary solutions of the Markovian evolution process.
The butterfly velocity $v_{\rm B}$ corresponds to the speed at which a domain wall between these two solutions propagates through the qudit array. For definiteness, we here will consider a domain wall separating a maximally random region to the left (smaller $x$) from a trivial region to the right (larger $x$). Such a domain wall propagates to the right (in the positive $x$ direction).

\subsection{Projected binary strings}
\label{subsec:3D}

Since the stochastic growth process defined by Eqs.\ (\ref{eq:gammasqavg})--(\ref{eq:Wgeneral}) refers to the basis of qudit operators --- the generalized Pauli matrices $\sigma_p$ ---, it has $q^2$ degrees of freedom per qudit.We may simplify the growth process by considering a projection of the classical probabilities $\rho_p(t)$ onto a ``binary'' probability string $\bar \rho_{\bar p}$, where the string index $\bar p$ is a list of binaries $[\bar p_0,\bar p_1,\ldots,\bar p_{L-1}]$ with $\bar p_x \in \{ \II, \XX \}$, where $\II$, $\XX$ represent a generalized Pauli matrix $\sigma_{p_x}$ with $p_x = 0$, $p_x \neq 0$, respectively. Writing $p \to \bar p$ to denote the binary string $\bar p$ corresponding to the Pauli string $p$, we define the binary-string distribution $\bar \rho_{\bar p}$ corresponding to  $\rho_p$ as
\begin{equation}
  \bar \rho_{\bar p} = \sum_{p \to \bar p} \rho_p.
  \label{eq:binarydef}
\end{equation}
(Note that the projection onto a binary string $\bar \rho_{\bar p}$ does not require that the underlying full Pauli-string weight $\rho_p$ has a binary form, in which it depends only on whether each generalized Pauli index $p_x = 0$ or $p_x \neq 0$. Nevertheless, in Sec.\ \ref{sec:5} we show that $\rho_p(t)$ relaxes to such a binary form after a finite time $\tau_{\rm b}$.)
The projected binary-string distribution corresponding to the maximally random steady-state solution (\ref{eq:rhomax}) is 
\begin{equation}
  \bar \rho^{\infty}_{\bar p}(t) = \prod_x \bar r^{\infty}_{\bar p_x},
\end{equation}
with
\begin{equation}
  \label{eq:rinfty}
  \bar r^{\infty}_{\bar p_x} = 
  \frac{1}{q^2} \left( \delta_{\bar p_x,\II} +
  (q^2-1) \delta_{\bar p_x,\XX} \right),
\end{equation}
whereas the projected binary-string distribution corresponding to the trivial solution (\ref{eq:rhotrivial}) is
\begin{equation}
  \label{eq:rhobinarytrivial}
  \bar \rho^0_{\bar p}(t) =
  \prod_{x} \delta_{\bar p_x,\II}.
\end{equation}
\begin{figure}[tb]
\centering
\includegraphics[scale=0.3, trim=1cm 2cm 2.5cm 0cm, clip]{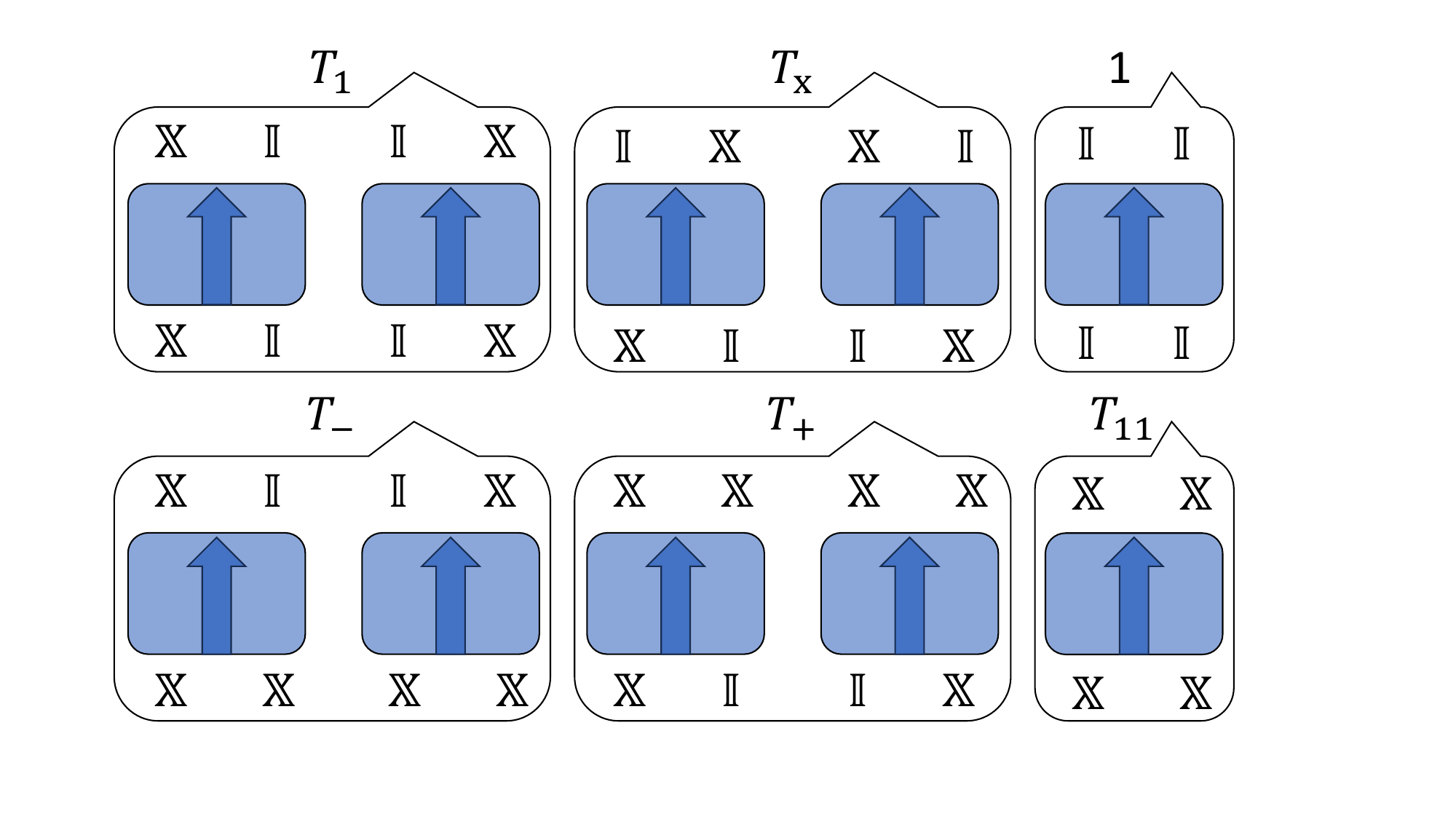} 
\caption{\justifying \small
Schematic showing the two-qudit gate processes corresponding to the rates $T_1$, $T_{11}$, $T_{\rm x}$, and $T_{\pm}$ defined in the main text. }
\label{fig:T}
\end{figure}

Using the properties of the function $\varphi(a,p)$, see App.\ \ref{app:1}, one finds that the Markovian evolution equation (\ref{eq:gammasqavg}) for the classical distribution $\rho_p$ gives a closed evolution equation for the projected binary distribution $\bar \rho_{\bar p}$,
\begin{equation}
  \bar \rho_{\bar p}(t) = \sum_{\bar a} \bar \rho_{\bar a}(t-1) T_{\bar a \bar p},
  \label{eq:rhobinaryevolution}
\end{equation}
with transition probabilities $T_{\bar a\bar p}$ that are products of pair-transition probabilities,
\begin{equation}
  T_{\bar a\bar p} =   
  \prod_{x\, {\rm even}}
  \left\{ \begin{array}{ll}
    T_{\bar a_x \bar a_{x+1};\bar p_x,\bar p_{x+1}} & \mbox{for $t$ even}, \\
    T_{\bar a_{x-1} \bar a_x;\bar p_{x-1} \bar p_x} & \mbox{for $t$ odd}.
  \end{array} \right.
\end{equation}
The nonzero pair transition rates $T_{\bar a_x,\bar a_y;\bar p_x,\bar p_y}$ for binary strings are $T_{\II \II;\II \II} = 1$, $T_{\II \XX;\II \XX} = T_{\XX \II;\XX \II} = T_1$, $T_{\II \XX;\XX \II} = T_{\XX \II;\II \XX} = T_{\rm x}$, $T_{\XX \XX;\II \XX} = T_{\XX \XX;\XX \II} = T_-$, $T_{\II \XX;\XX \XX} = T_{\XX \II;\XX \XX} = T_+$, and $T_{\XX \XX;\XX \XX} = T_{11}$. Here $T_{1}$ and $T_{11}$ describe processes at the two-qudit gate for which the (binary) qudit states are preserved, $T_{\rm x}$ describes a swap between zero and nonzero qudit states, and $T_{\pm}$ describe processes in which a zero qudit is turned into a nonzero one or vice versa, see Fig.\ \ref{fig:T}. From Eq.\ (\ref{eq:Wgeneral}) we obtain
\begin{align}
  \label{eq:T}
  T_1 =&\, \frac{1}{q^2+1} \left( 1 + \frac{q^2(\mbox{Re}\, A_1 + A_2 + A_3)}{q^4-1} \right), \nonumber \\
  T_+ =&\, (q^2-1) T_- \nonumber \\ =&\,
  \frac{q^2-1}{q^2+1} \left( 1 + \frac{\mbox{Re}\, A_1 q^2 - A_2 - A_3}{q^4-1} \right).
\end{align}
The remaining transition rates can be obtained from the probability conservation rules 
\begin{align}
  1 =&\,  T_1 + T_{\rm x} + T_+ \nonumber \\ =&\, 2 T_- + T_{11}.
  \label{eq:Tcons}
\end{align}
The reduction to a stochastic growth model of binary strings $\rho_p(t)$ allows for easy numerical simulations even in the limit of large $q$. It also allows for an efficient approximate solution, as we show in the next Subsections.
\subsection{Right-propagating $n$-point density}
\label{sec:3e}
We say that the binary string $\bar p$ ``ends at $x$'' if $x$ is the rightmost qudit position with $\bar p_x = \XX$. (This implies that $\bar p_y = \II$ for all $y > x$.) Following Ref.\ \cite{nahum_operator_2018}, we define the ``right-propagating density'' $\bar \rho_{\rm }^{(0)}(\Delta x;t)$ as the fraction of Pauli strings ending at $x = t + \Delta x$,
\begin{equation}
  \bar \rho_{\rm }^{(0)}(\Delta x;t) = \sum_{\textrm{$\bar p$ ends at $t + \Delta x$}} \bar \rho_{\bar p}(t).
\end{equation}
(Note that the argument $\Delta x$ is measured with respect to the ballistic propagation at unit velocity.)
We also define the ``right-propagating $n$-point density''
\begin{widetext}
\begin{equation}
  \label{eq:rhoRdef}
    \bar\rho_{\rm }^{(n)}(\Delta x;t;\bar a_n,\ldots,\bar a_1)
 =\,
  \sum_{\textrm{$p$ ends at $t + \Delta x$}} \bar \rho_{\bar p}(t)\,
  \delta_{\bar p_{t+\Delta x-1},\bar a_{1}} \ldots \delta_{\bar p_{t+\Delta x-n},\bar a_{n}},
\end{equation}
which is the fraction of Pauli strings that end at $t + \Delta x$ and that have the sequence $\bar a_n,\ldots,\bar a_1,\XX$ as their rightmost nontrivial entries, whereby the rightmost ``$\XX$'' appears at site $t + \Delta x$.

From the stochastic evolution equation for the full binary probability density $\bar \rho_{\bar p}(t)$ we may deduce evolution equations for the right-propagating density and for the right-propagating $n$-point densities. These are
\begin{align}
  \label{eq:rhoevol1}
  \bar \rho_{\rm }^{(n)}(\Delta x;t;\bar p_n,\ldots,\bar p_1) =&\,
  \sum_{\bar a_1,\ldots,\bar a_{n}} 
  T_{\bar a_{n} \bar a_{n-1};\bar p_{n} \bar p_{n-1}} \ldots T_{\bar a_2,\bar a_1;\bar p_2,\bar p_1}
  T_{\XX\II;\XX\II} \bar \rho_{\rm }^{(n)}(\Delta x+1;t-1;\bar a_n,\ldots,\bar a_1)
  \\ \nonumber &\, \mbox{}
  + \sum_{\bar a_1,\ldots,\bar a_{n+1}}
  T_{\bar a_{n+1}\bar a_{n};\bar p_{n}\bar p_{n-1}}
  \ldots T_{\bar a_3,\bar a_2;\bar p_2,\bar p_1}
  T_{\bar a_1 \XX;\XX\II}    \bar \rho_{\rm }^{(n+1)}(\Delta x+2;t-1;\bar a_{n+1},\bar a_n,\ldots,\bar a_{1}) 
\end{align}
if $n$ and $\Delta x$ are both even. The summation variables $\bar a_j$ take the values $\II$, $\XX$, $j=1,\ldots,n$. In the same manner, we find
\begin{align}
  \label{eq:rhoevol2}
  \bar \rho_{\rm }^{(n)}(\Delta x;t;\bar p_n,\ldots,\bar p_1) =&\,
  \sum_{\bar a_1,\ldots,\bar a_{n}} \sum_{\bar p_{n+1}} T_{\bar a_{n}\bar a_{n-1};\bar p_{n+1}\bar p_{n}} \ldots T_{\bar a_2,\bar a_1;\bar p_3,\bar p_2} T_{\XX\II;\bar p_1 \XX}  \bar \rho_{\rm }^{(n)}(\Delta x;t-1;\bar a_{n},\ldots,\bar a_{1})
   \\ \nonumber &\, \mbox{}
  +
  \sum_{\bar a_1,\ldots,\bar a_{n+1}} \sum_{\bar p_{n+1}} T_{\bar a_{n+1}\bar a_{n};\bar p_{n+1}\bar p_{n}} \ldots T_{\bar a_3,\bar a_2;\bar p_3,\bar p_2} T_{\bar a_1 \XX;\bar p_1 \XX}
   \bar \rho_{\rm }^{(n+1)}(\Delta x+1;t-1;\bar a_{n+1},\bar a_n,\ldots,\bar a_1)
  , 
\end{align}
\end{widetext}
if $n$ is even and $\Delta x$ is odd. 
To avoid spurious even-odd effects, we will not consider the evolution equations for the right-propagating $n$-point density with $n$ odd.

The evolution equations (\ref{eq:rhoevol1}) and (\ref{eq:rhoevol2}) for the right-propagating $n$-point density both involve the $(n+1)$-point density. For Haar-distributed two-qudit gates, the transition probabilities are such, that the $(n+1)$-point density appears in the combination $\sum_{\bar a_{n+1}} \bar \rho^{(n+1)}(\Delta x,t;\bar a_{n+1},\bar a_{n},\ldots,\bar a_1) = \bar \rho^{(n)}(\Delta x,t;\bar a_{n},\ldots,\bar a_1)$, so that the evolution equations (\ref{eq:rhoevol1}) and (\ref{eq:rhoevol2}) can be closed \cite{nahum_operator_2018}. In the same way, we can get closed-form evolution equations for the right-propagating densities for generic unitary-invariant two-qudit gate distributions in the limit of large $q$, because the difference between $T_{-}$ and $T_{\rm x}$ vanishes in this limit. Such simplification does not occur at finite $q$. To arrive at a closed set of equations in this case, we truncate the evolutions (\ref{eq:rhoevol1}) and (\ref{eq:rhoevol2}) at sufficiently high order $n$ by replacing the $(n+1)$-point density by its maximally random approximation,
\begin{align}
  \label{eq:rhonn}
  \lefteqn{ \bar \rho^{(n+1)}_{\rm }(\Delta x;t;\bar p_{n+1},\bar p_{n},\ldots,\bar p_1)}
  ~~~~~~~~~~~~ \nonumber \\ =&\,
  \bar \rho^{(n)}_{\rm }(\Delta x;t;\bar p_n,\ldots,\bar p_1) \bar r^{\infty}_{\bar p_{n+1}},
\end{align}
where $\bar r^{\infty}_{\bar p_x}$ was defined in Eq.\ (\ref{eq:rinfty}). We expect that such truncation is justified for $n \gg n_{\rm DW}$, where $n_{\rm DW}$ is the width of the domain wall between the trivial and maximum-entropy regions of the Pauli-string density $\rho_p$, see Eqs.\ (\ref{eq:rhomax}) and (\ref{eq:rhotrivial}).
Application of this truncation procedure to the right-moving density $\bar \rho^{(0)}(x;t)$ gives the evolution equation
\begin{align}
  \label{eq:rhoeven}
  \bar \rho_{\rm }^{(0)}(\Delta x;t) =&\,
  T_{1} \bar \rho_{\rm }^{(0)}(\Delta x+1;t-1)
  \\ \nonumber &\, \mbox{} +
  \frac{ T_{\rm x} + (q^2-1) T_{-}}{q^2}
     \bar \rho_{\rm }^{(0)}(\Delta x+2;t-1)
\end{align}
if $\Delta x$ is even and
\begin{align}
  \label{eq:rhoodd}
  \lefteqn{
  \bar \rho_{\rm }^{(0)}(\Delta x;t) = 
(T_{\rm x} + T_{+}) \bar \rho_{\rm }^{(0)}(\Delta x;t-1)} 
  \\ \nonumber &\, \mbox{} +
  \frac{T_1 + T_+ + (q^2-1)(T_{11} + T_{-})}{q^2}
  \bar \rho_{\rm }^{(0)}(\Delta x+1;t-1)
\end{align}
if $\Delta x$ is odd, where we used Eqs.\ (\ref{eq:T}) and (\ref{eq:Tcons}) for the binary-string transition probabilities.

\subsection{Drift-Diffusion process}

We now describe, how the evolution equations (\ref{eq:rhoevol1}) and (\ref{eq:rhoevol2}) with the cut-off procedure (\ref{eq:rhonn}) can be mapped to a drift-diffusion process.
To simplify the notation, we combine $\bar \rho^{(n)}_{\rm }(x;t)$ and $\bar \rho^{(n)}_{\rm }(x+1,t)$ into a two-component spinor,
\begin{align}
  \label{eq:Rdef}
  R_{\rm }^{(n)}(\Delta x;t) = \begin{pmatrix} \bar \rho_{\rm }^{(n)}(\Delta x;t) \\
  \bar \rho_{\rm }^{(n)}(\Delta x + 1;t) \end{pmatrix},
\end{align}
where now $\Delta x$ is always even.
The evolution equations (\ref{eq:rhoevol1})--(\ref{eq:rhoevol2}) with the truncation prescription (\ref{eq:rhonn}), may then be represented as
\begin{align}
  R_{\rm }^{(n)}(\Delta x;t) =&\,
  D^{(n)} R_{\rm }^{(n)}(\Delta x;t-1)  \nonumber \\ &\, \mbox{}
  + D'^{(n)} R_{\rm }^{(n)}(\Delta x+2;t-1),
  \label{eq:Revol}
\end{align}
where $D^{(n)}$ and $D'^{(n)}$ are $2^{n+1} \times 2^{n+1}$ matrices. In the special case $n=0$, see Eqs.\ (\ref{eq:rhoeven}) and (\ref{eq:rhoodd}), the matrices $D^{(0)}$ and $D'^{(0)}$ read
\begin{align}
  D^{(0)} =&\, \begin{pmatrix}
  0 & T_1 \\
  0 & T_{\rm x} + T_+ \end{pmatrix}, \nonumber \\
  D'^{(0)} =&\, \begin{pmatrix}
  \frac{T_{\rm x} + (q^2-1) T_-}{q^2} & 0 \\
  \frac{T_1 + T_+ + (q^2-1)(T_{11} + T_-)}{q^2} & 0 \end{pmatrix}.
\end{align}
The transition matrices $D^{(n)}$ and $D'^{(n)}$ satisfy the normalization rule
\begin{equation}
  \sum_{i} (D^{(n)}_{ij} + D'^{(n)}_{ij}) = 1
  \label{eq:Dsum}
\end{equation}
for each $j=1,\ldots,2^{n+1}$. We denote the left-eigenvectors, right-eigenvectors, and eigenvalues of $D^{(n)} + D'^{(n)}$ by $\tilde V_j^{(n)}$, $V_j^{(n)}$, and $d_{j}^{(n)}$, respectively. Equation (\ref{eq:Dsum}) ensures that the largest eigenvalue $d_1^{(n)} = 1$, with left-eigenvector $\tilde V_1^{(n)} = (1,1,\ldots,1)^{\rm T}$. 

In the long-time limit, the solution of Eq.\ (\ref{eq:Revol}) is of the form
\begin{equation}
  R^{(n)}_{\rm }(\Delta x,t) = R_{{\rm }1}^{(n)}(\Delta x,t) V_1^{(n)},
  \label{eq:Rntlong}
\end{equation}
where $R_{{\rm }1}^{(n)}(\Delta x,t)$ satisfies the drift-diffusion equation (see App.\ \ref{app:5} for details)
\begin{align}
  \label{eq:driftdiffusion}
  \partial_t R_{{\rm }1}^{(n)}(\Delta x,t) =&\,
  (1-v_{\rm B}^{(n)} )
  \partial_{\Delta x} R_{{\rm }1}^{(n)}(\Delta x,t)
  \nonumber \\ &\, \mbox{}
  + \frac{{\cal D}^{(n)}}{2} \partial_{\Delta x}^2 
  R_{{\rm }1}^{(n)}(\Delta x,t).
\end{align}
Here
\begin{align}
  v_{\rm B}^{(n)} =&\, 1 - 2 d_{11}'^{(n)}
  \label{eq:vBn} 
\end{align}
is the butterfly velocity and
\begin{align}
  {\cal D}^{(n)} =&\, 4 d_{11}'^{(n)}[1 -  d_{11}'^{(n)}]
  + 8 \sum_{j \neq 1} \frac{d'^{(n)}_{1j} d'^{(n)}_{j1}}{1 - d_j^{(n)}}
  \label{eq:DDn}
\end{align}
the diffusion constant governing the diffusive spreading of the front. (Recall that $\Delta x$ is measured with respect to a ballistic propagation at unit velocity so that the drift velocity measured with respect to $\Delta x$ is $v_{\rm B} - 1$.) In Eqs.\ (\ref{eq:vBn}) and (\ref{eq:DDn}) we abbreviated 
\begin{equation}
  d_{ij}'^{(n)} = \tilde V_i^{(n){\rm T}} D'^{(n)} V_j^{(n)}.
  \label{eq:dijdef}
\end{equation}
In Eq.\ (\ref{eq:Rntlong}), the function $R_1^{(n)}(\Delta x,t)$ describes the propagation of the ends of the Pauli strings, whereas the right-eigenvector $V_1^{(n)}$ contains information on the structure of the Pauli strings immediately behind the end. Appendix \ref{app:5} also contains an expression for the diffusion constant ${\cal D}^{(n)}$ for the case that the sum $D^{(n)} + D'^{(n)}$ of the transition matrices is not diagonalizable so that no complete basis of left- and right-eigenvectors exists.

For the lowest-order approximation $n=0$, the left- and right-eigenvectors of $D^{(0)} + D'^{(0)}$ can be found in closed form and we obtain
\begin{align}
  \label{eq:vB0result}
  v_{\rm B}^{(0)} =&\,  \frac{(q^2-1)(1-T_1)}{q^2-1 + T_1 (q^2+1)}, \\
  \label{eq:DD0result}
  {\cal D}^{(0)} =&\, \frac{4(q^2 + 1)(q^2 - 1 + T_1)T_1 (1-T_1)}{[q^2-1 + (q^2+1) T_1]^2}.
\end{align}
For Haar-distributed two-qubit operators, the lowest-order approximation is exact and Eqs.\ (\ref{eq:vB0result}) and (\ref{eq:DD0result}) reproduce the results (\ref{eq:vBHaar}) and (\ref{eq:DHaar}) quoted in the introduction
\cite{nahum_operator_2018}. For a trivial circuit, one has $v_{\rm B} = 0$ and ${\cal D} = 0$.

\section{Butterfly velocity for Poisson kernel}
\label{sec:4}

As an application, we now consider a random unitary circuit for which the two-qudit gate operators ${\cal U}_{x,x+1}$ are taken from the Poisson-kernel distribution (\ref{eq:Poisson}) with ensemble average $\langle {\cal U}_{x,x+1} \rangle = \alpha \openone$. The Poisson kernel interpolates between the Haar distribution for $\alpha = 0$ and a trivial circuit with no information spreading for $|\alpha| = 1$.\textcolor{black}{ An alternative interpolating random matrix ensemble for the two-qudit evolution operators can be obtained from the Dyson's Brownian motion ensemble of random matrices \cite{Dyson1962Brownian,Dyson1972Brownian,tang2024brownian}. This interpolating ensemble is discussed in App.\ \ref{app:Brownian}.}

\begin{figure}[tb]
\centering
\includegraphics[trim=.5cm 0cm 1cm 0.1cm, clip,scale=0.55]{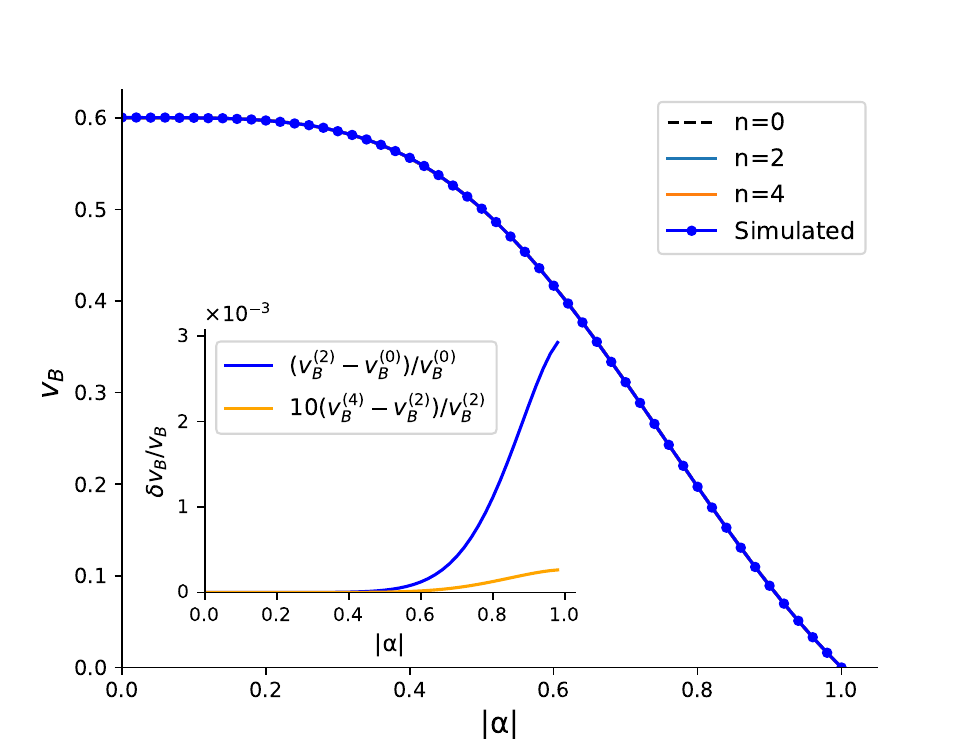} 
\label{fig:subfig-a}    
\includegraphics[trim=.5cm 0cm 1.5cm 0.1cm, clip,scale=0.6]{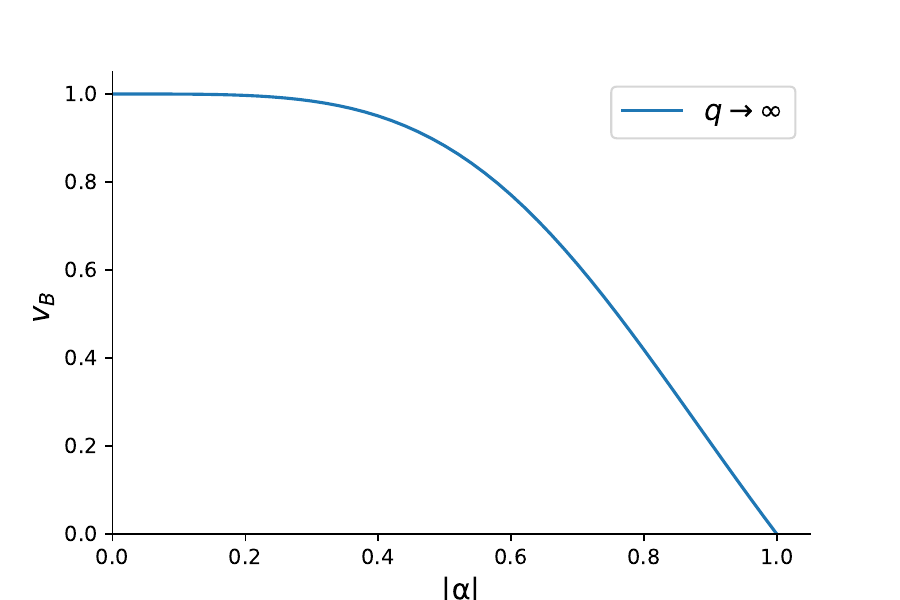} 
 \label{fig:subfig-b}
\caption{\justifying
\small Butterfly velocity for a random unitary circuit with Poisson kernel-distributed two-qudit gates for $q=2$ and $q \to \infty$. The interpolation parameter $\alpha$ determines the ensemble average $\langle {\cal U}_{x,x+1} \rangle = \alpha \openone$ of the two-qudit gate operator ${\cal U}_{x,x+1}$. The solid curves show the approximations $v_{\rm B}^{(n)}$ for order $n=0$, $2$, and $4$. The inset illustrates the relative differences between the fourth-order and second-order approximations, as well as the relative differences between the second-order and zeroth-order approximations. For $q\to \infty$ the approximation scheme is exact already for $n=0$. The data points are from a direct numerical simulation of the stochastic process (\ref{eq:rhobinaryevolution}) for $L = 320$, which is based on $N = 2\times 10^5$ independent realizations. \textcolor{black}{ Due to the rapid convergence of the truncation scheme, the analytical results for orders $n=0$, $2$, and $4$ and the numerical simulation data in the top panel are indistinguishable. }}
\label{fig:Poisson} 
\end{figure}

\begin{figure}[tb]
\centering
\includegraphics[trim=0.8cm 0cm 1.9cm 0.1cm, clip,scale=0.56]{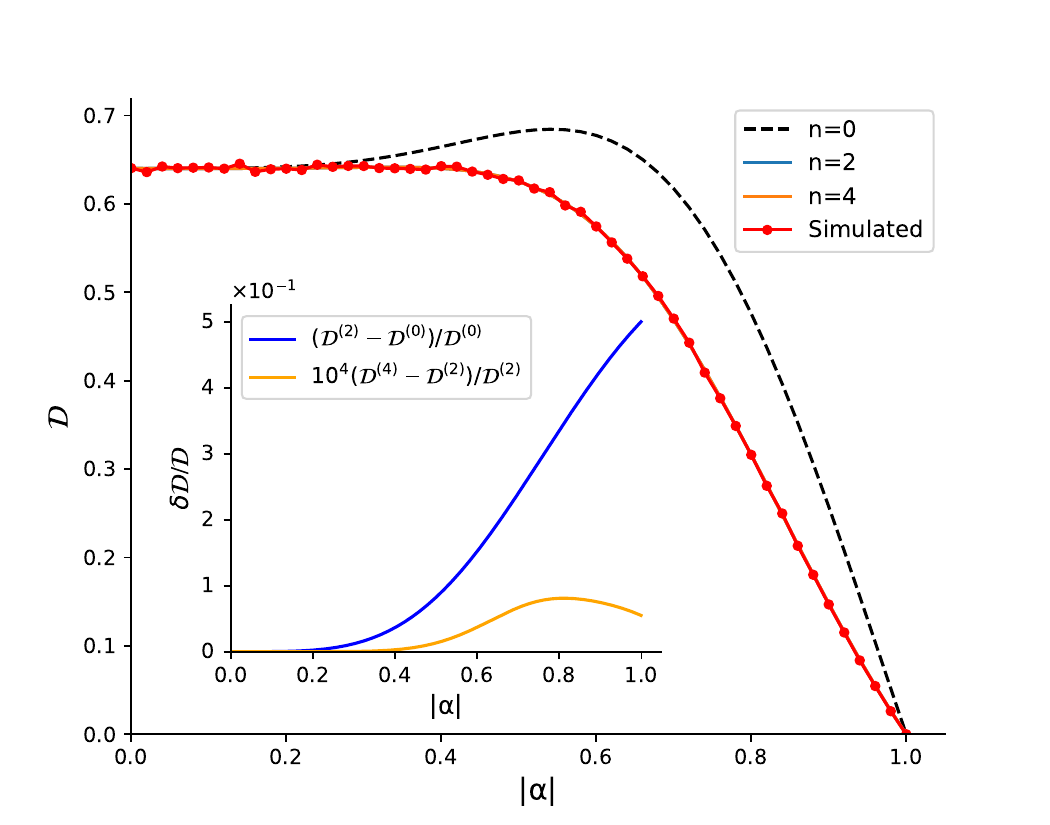}
\label{sub1} 
\vfill
\includegraphics[trim=.5cm 0cm 1.5cm 0.1cm, clip,scale=0.58]{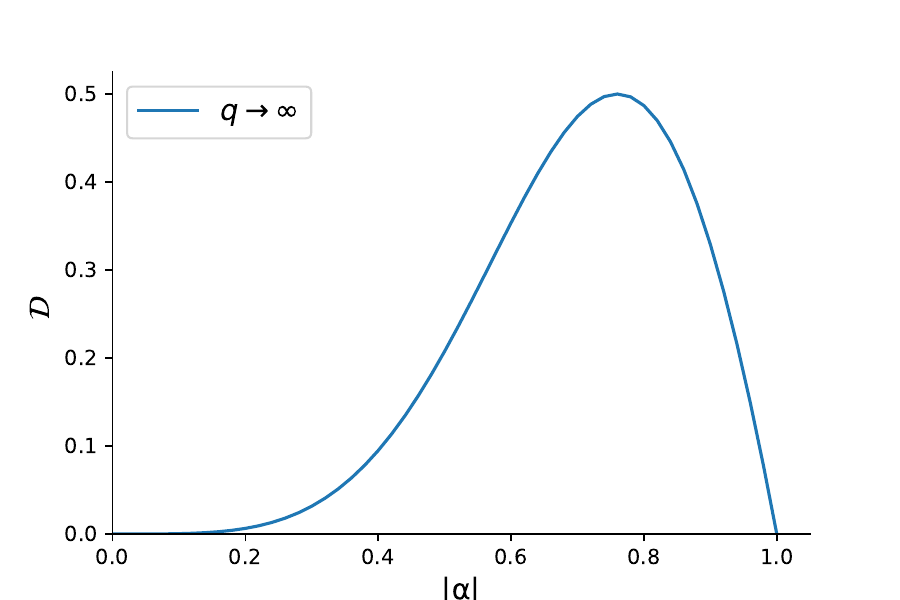}
\label{sub2}
     
\caption{\justifying \small Same as Fig.\ \ref{fig:Poisson}, but for the diffusion constant ${\cal D}$ for a random unitary circuit with Poisson kernel-distributed two-qudit gates. The small fluctuations of the data points around the theoretical curve are attributed to the remaining statistical fluctuations of the large, but finite number of realizations $N$ of the stochastic growth process in the numerical simulations. \textcolor{black}{ Due to the rapid convergence of the truncation scheme, the analytical results for orders $n=$ $2$ and $4$ in the top panel are indistinguishable. }}
     \label{fig:PoissonD}
 \end{figure}

For the Poisson kernel, the coefficients $A_1$, $A_2$, and $A_3$ in Eq.\ (\ref{eq:Wgeneral}) are (see App. \ref{app: 8})
\begin{align}
  \label{eq:APoisson1}
  A_1 =&\, \frac{8 |\alpha|^4}{q^4-9}
  - 2 \frac{(q^2-1) |\alpha|^{2 q^2-2}}{q^2(q^2-3)}
  + 2 \frac{(q^2+1) |\alpha|^{2 q^2+2}}{q^2(q^2+3)}, 
\end{align}
\begin{align}
  A_2 =&\, \frac{6 |\alpha|^4 (q^4 + 1)}{(q^4-4)(q^4-9)}
  - \frac{(q^2-1) |\alpha|^{2 q^2-2}}{q^2-3}
  \nonumber \\ &\, \mbox{}
  + 2 \frac{(q^4-1) |\alpha|^{2 q^2}}{q^4-4}
  - \frac{(q^2+1) |\alpha|^{2 q^2+2}}{q^2+3},
\label{eq:APoisson2}    
\end{align}
\begin{align}
  \label{eq:APoisson3}
  A_3 =&\,
  \frac{(6 + 15 q^4 - 10 q^8 + q^{12})|\alpha|^4}{(q^4-4)(q^4-9)}
  - \frac{(q^2-1) |\alpha|^{2 q^2-2}}{q^2-3}
  \nonumber \\ &\, \mbox{}
  - 2 \frac{(q^4-1) |\alpha|^{2 q^2}}{q^4-4}
  - \frac{(q^2+1) |\alpha|^{2 q^2+2}}{q^2+3}.
\end{align}
One verifies that these expressions reproduce the limits $A_1 = A_2 = A_3 = 0$ corresponding to the Haar-distributed two-qubit gate operators for $\alpha = 0$ and $A_1 = A_2 = 0$, $A_3= q^4-1$ for the limit $\alpha \to 1$ corresponding to trivial gate operators.
For the butterfly velocity and the diffusion constant, we then arrive at the expressions
\begin{align}
  v_{\rm B}^{(0)} =&\frac{q^2-1}{q^2+1}\frac{a-b}{a+b}, \\
  {\cal D}^{(0)} =&\frac{4}{(q^2+1)^2}\frac{(a-b)(b q^2 + a)(b+a q^2)}{a(a+b)^2}\, , 
\end{align}
with                    \begin{align}
  a=&\, q^6 - 9 q^2 \nonumber \\
 b =&\, (q^6-5q^2)|\alpha|^4-(6+2q^2)|\alpha|^{2q^2-2} \nonumber \\ &\, \mbox{} +(6-2q^2)|\alpha|^{2+2q^2}\, . 
\end{align}
In the large-$q$ limit (while keeping $\alpha$ fixed), these results simplify to
\begin{align}
  v_{\rm B}^{(0)} =&\, \frac{1 - |\alpha|^4}{1 + |\alpha|^4}, \\
  {\cal D}^{(0)} =&\, \frac{4 |\alpha|^4 (1 - |\alpha|^4)}{(1 + |\alpha|^4)^2}.
\end{align}
Whereas the diffusing front is sharp in the limit $q \to \infty$ for the Haar-distributed case $\alpha = 0$, the diffusion constant is nonzero in this limit for $|\alpha| > 0$.

\textcolor{black}{ For $|\alpha|$ close to $1$, the probability that the front moves in a single time step becomes small. In this limit, the discrete time evolution of the random circuit can be approximated by a continuous random process. Such a continuous random process is again characterized by a drift velocity and diffusion constant \cite{xu_locality_2019,xu_scrambling_2022}. To find these within our approach, we set $|\alpha|^2 = e^{-\lambda}$, rescale $t' = \lambda t$, and take the limit $\lambda \to 0$. Using $t'$ as the time variable for the random continuous process, the lowest-order approximations for the drift velocity and diffusion constant become
\begin{align}
  v_{\rm B}'^{(0)} =&\, \frac{(q^2-1)^2(q^2+2)}{q^2(q^2+1)(q^2+3)}, \label{eq:vBcontinuous} \\
  {\cal D}'^{(0)} =&\, \frac{2(q^4+q^2-2)}{q^2(q^2+3)} \label{eq:Dcontinuous}.
\end{align}}

Figures \ref{fig:Poisson} and \ref{fig:PoissonD} show the zeroth-order results for the butterfly velocity vs.\ $|\alpha|$ for $q=2$ and for the limit $q \to \infty$, together with a numerical evaluation of the $n$th order approximations $v_{\rm B}^{(n)}$ and ${\cal D}^{(n)}$ for $n=2$ and $n=4$. The data points in Figs.\ \ref{fig:Poisson} and \ref{fig:PoissonD} show the result of a direct numerical simulation of the binary stochastic evolution described by Eq.\ (\ref{eq:rhobinaryevolution}). As depicted in Fig\ \ref{fig:Poisson}, already the zeroth-order approximation is extremely accurate for the butterfly velocity.\textcolor{black}{ This includes the limit $|\alpha| \to 1$, in which the butterfly velocity goes to zero and the random circuit becomes a random continuous process. }For the diffusion constant, there is a slightly larger difference between $n=0$ and $n=2$, but here, too, the difference between the approximations at orders $n=2$ and $n=4$ is no longer visible. Hence, for both butterfly velocity and diffusion constant, we conclude that our approximation scheme converges within a distance of only a few sites from the right end of the Pauli string, which suggests that the width $n_{\rm DW}$ of the domain wall between the trivial and maximum-entropy regions of the Pauli string distribution is of that same order.

\begin{figure}
    \centering
    \includegraphics[trim=0cm 0cm 0cm 0cm, clip,scale=0.6]{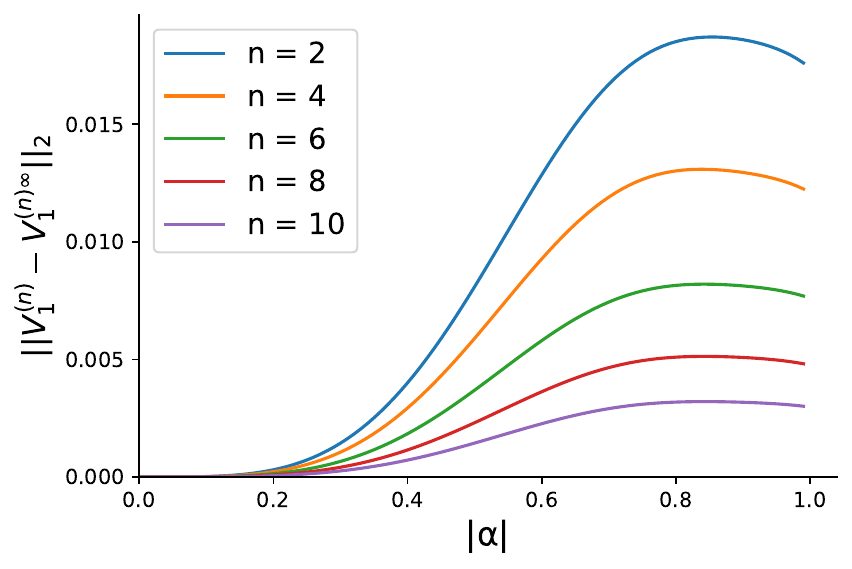}
    \caption{\justifying \small
    The norm of the difference between $V_1^{(n)}$ and $\bar{V}_1^{(n)}$ for a random unitary circuit with Poisson-kernel-distributed two-qudit gates, plotted as a function of $|\alpha|$ for $n = 2, 4, 6, 8, 10$ with $q = 2$.}
    \label{fig:nW}
\end{figure}

\textcolor{black}{ Sufficiently far from the moving front, the maximally random approximation of Eq.\ (\ref{eq:rhonn}) should hold for the $n$-point density $\rho^{(n)}(\Delta x;t;\bar a_n,\ldots,\bar a_1)$. In the long-time limit, the information on the dependence of the $n$-point density on the Pauli string elements $\bar a_n$, \ldots, $\bar a_1$ is encoded in the right-eigenvector $V_1^{(n)}$ of the transition matrix $D^{(n)} + D'^{(n)}$, see Eq.\ (\ref{eq:Rntlong}). }To further verify the validity of our approximation scheme, we compute the difference between the right-eigenvectors $V_1^{(n)}$ of the transition matrix $D^{(n)} + D'^{(n)}$ relevant for the information spreading at large times and its approximation according to the {\em Ansatz} (\ref{eq:rhonn}). Hereto, we define
\begin{align}
  V_1^{(n) \infty}(\bar a_{n},\bar a_{n-1},\ldots,\bar a_{1}) =&\,
  r^{\infty}_{\bar a_{n}}r^{\infty}_{\bar a_{n-1}}  \\ \nonumber &\, \mbox{} \times
  V_1^{(n-2)}(\bar a_{n-2},\ldots,\bar a_{1}).
\end{align}
\textcolor{black}{(We compare $V^{(n)}$ and $V^{(n-2)}$ to avoid even-odd effects intrinsic to the brickwork structure of the random circuit.)} In Fig.\ \ref{fig:nW} we show the two-norm $\|V_1^{(n)} - {V}_1^{(n)\infty}\|_2$ vs.\ the truncation order $n$. The figure shows that the norm exhibits rapid convergence, particularly when the ensemble closely approximates the Haar measure. 

\section{Full distribution $\rho_p$ and out-of-time-ordered correlator}
\label{sec:5}

\subsection{Convergence of $\rho_p$ to a binary distribution}
\label{sec:5a}

\begin{figure}
    \centering
\includegraphics[width=1\linewidth]{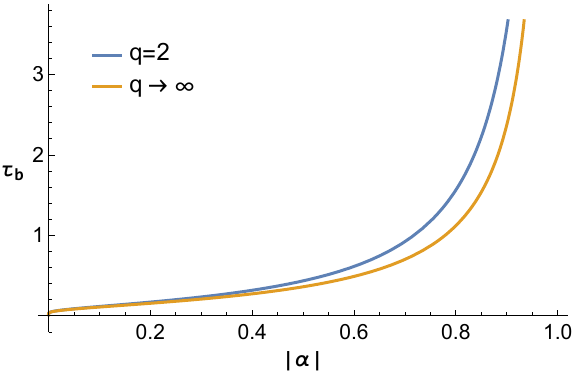}
    \caption{\justifying \small Time scale $\tau_{\rm b}$ for the approach to a binary distribution for a random unitary circuit with Poisson-kernel distribution for the two-qudit gates as a function of $|\alpha|$ for $q = 2$ and for the limit $q \to \infty$. In the Haar limit $\alpha = 0$, a binary distribution is reached after a single time step.}
    \label{fig:decay}
\vspace{-0.6cm}
\end{figure}

We have seen that the butterfly velocity and the diffusion constant could be obtained by considering the projected binary strings $\bar \rho_{\bar p}$ with two degrees of freedom per qudit. We now return to the full distribution $\rho_{p}$, which has $q^2$ degrees of freedom per qubit. Below, we show that in the long-time limit, the full Pauli-string weight $\rho_p$ approaches a binary form, in which $\rho_p$ depends only on whether the Pauli string entries $p_x$ are $0$ or different from $0$. Specifically, we show that in the long-time limit $\rho_p$ approaches $B \rho_p$, with
\begin{align}
  B \rho_{p} = \left( \prod_x B_{p_x,\bar p_x} \right) \bar \rho_{\bar p},
\end{align}
where $\bar p$ is the binary string corresponding to $p$, $\bar \rho_{\bar p}$ the binary-string distribution corresponding to $\rho_p$, and
\begin{equation}
  B_{p_x,\bar p_x} = 
  \delta_{p_x,0} \delta_{\bar p_x,\II} + 
  \frac{(1 - \delta_{p_x,0}) \delta_{\bar p_x,\XX}}{q^2-1}.
\end{equation}
We refer to the distribution $B \rho_{p}$ as the ``binary distribution corresponding to $\rho_{p}$''.
In App.\ \ref{app:3} we verify that if a probability distribution $\rho_p(t)$ is of binary type, {\em i.e.}, $\rho_p(t) = B \rho_p(t)$, this property is preserved under time evolution. To prove that a distribution $\rho_p(t)$ approaches $B \rho_p(t)$ for large times, we consider the two-norm of the difference
\begin{equation}
  || \rho(t) - B \rho(t) ||_2 =
\sqrt{\sum_{p}|\rho_p(t) - B \rho_p(t)|^2}.
\end{equation}
In a single time step, it is bounded by
\begin{equation}
  || \rho(t+1) - B \rho(t+1) ||_2 \le
  ||W||'_{\infty}
  || \rho(t) - B \rho(t) ||_2,
\end{equation}
where $||W||'_{\infty}$ is the operator norm of the transition probability matrix $\langle |W_{ap}|^2 \rangle$ of Eq.\ (\ref{eq:gammasqavg}) --- i.e., the largest singular value ---, after exclusion of the binary Pauli-string weights for which $\rho_p - B \rho_p = 0$. Per qudit there are two Pauli-string weights with $\rho_p = B \rho_p$, the maximally random distribution and the trivial one. The singular values of $\langle |W_{ap}|^2 \rangle$ are $1$, $\lambda_{\pm}$, and $\lambda_{\rm i}$, with
\begin{align}
  \lambda_{\pm} =&\, \frac{A_2 + A_3 \pm q^2 \mbox{Re}\, A_1}{q^4-1}, \nonumber \\
  \lambda_{\rm i} =&\, \left| \frac{A_3 - A_2 + i q^2 \mbox{Im}\, A_1}{q^4-1}\right|.
\end{align}
The degeneracies of these singular values are $N_1 = 2$, $N_{\pm} = 3 \mp q^2/2+q^4/4$, $N_{\rm i} = q^2/-8$ if $q$ is even and $N_1 = 2$, $N_{\pm} = -3/4 \mp q^2/2 + q^4/4$, $N_{\rm i} = (q^4-3)/2$ if $q$ is odd.
The singular value $1$ belongs to the two binary weights. Hence, the largest singular value of the two-qudit transition probability matrix after exclusion of the binary Pauli-string weights is
\begin{align}
  \lambda_W =&\, \max(\lambda_+,\lambda_{\rm i}).
\end{align}
Since $\lambda_W < 1$ except for the trivial circuit, it follows that the probability distribution $\rho_p(t)$ exponentially approaches the corresponding maximum entropy distribution $B \rho_p(t)$. For a Haar-random circuit, one has $\lambda_W = 0$, so that the distribution $\rho_p$ is binary already after a single time step.
For a circuit with $\lambda_{\rm W} > 0$ we can define the decay time 
\begin{equation}
  \tau_{\rm b} = \frac{-1}{\ln \lambda_W}
\end{equation}
to characterize the time scale to converge to a binary distribution. Figure \ref{fig:decay} shows $\tau_{\rm b}$ for a random unitary circuit with Poisson-kernel distribution for the two-qudit gates. 
In the limit $\alpha \to 1$, the asymptotic expression for the decay time scale is
\begin{equation}
    \tau_{\rm b}\approx
      \frac{q^2+1}{q^2-1}\frac{1}{4(1-\alpha)}-\frac{5q^2-3}{8(q^2-1)}.
\end{equation}
(Here the limit $\alpha \to 1$ is taken at fixed $q$. If one first sends $q \to \infty$ and then takes the limit $\alpha \to 1$, one has $\tau_{\rm b} \approx (1/4)(1-\alpha)^{-1} - 1/8$.)

\subsection {Relation to out-of-time-order correlator}

In the literature, operator spreading is also described via the ``out-of-time-ordered correlator'' (OTOC) $C(x,y;t)$ \cite{nahum_operator_2018,von_keyserlingk_operator_2018,xu_scrambling_2022}, which is defined as
\begin{equation}
  C(x,y,t)= \frac{1}{2 q^{L}} \tr \left|\left[O_x(t),O_y(0)\right]\right|^2.
  \label{eq:OTOC}
\end{equation}
Here $O_x \equiv O_x(0)$ and $O_y \equiv O_y(0)$ are Pauli-string operators that contain nontrivial generalized Pauli matrices $\sigma_{p_x}$ and $\sigma_{p_y}$ at qudits $x$ and $y$, respectively, while being trivial at all other positions. Using Eq.\ (\ref{eq:Ot}) for $O_x(t)$, we may express the average $\langle C(x,y;t) \rangle$ over realizations of the random circuit as
\begin{align}
  \langle C(x,y,t) \rangle =&\,  1- q^{-L} 
  \Re \langle \tr O_x(t)O_y(0) O_x(t)^\dagger O_y(0)^\dagger \rangle \nonumber \\
  =&\, 1 - q^{-1} \sum_{a} \rho_a(t) \Re \tr \sigma_{a_y} \sigma_{p_y}
  \sigma_{a_y}^{\dagger} \sigma_{p_y}^{\dagger},
  \label{eq: otoc1}
\end{align}
where $\rho_p(t) = \langle |\gamma_p(t)|^2 \rangle$ is the classical Pauli-string probability distribution for the operator $O_x(t)$. Since the ensemble-averaged OTOC (\ref{eq:OTOC}) depends on the distance $x-y$ only, we denote it by $\langle C(x-y,t) \rangle$ henceforth.

In the previous Subsection, we have shown that $\rho_a(t)$ approaches the binary form for times $t \gg \tau_{\rm b}$, so that we can express $\langle C(s,t) \rangle$ in terms of the projected binary-string distribution $\bar \rho_{\bar a}(t)$ \cite{xu_scrambling_2022},
\begin{align}
 C(s,t) =&\, 1-
  \sum_{\bar a} \left( \delta_{\bar a_s,\II} -
  \frac{\delta_{\bar a_s,\XX}}{q^2-1} \right)
  \bar \rho_{\bar a}(t).
  \label{eq:Cxyt1}
\end{align}
Considering, without loss of generality, the case $s \ge 0$, we may assume that the contribution of Pauli strings $\bar a$ with a left end at position $\ge x$ to the summation in Eq.\ (\ref{eq:Cxyt1}) is exponentially small, so that it is sufficient to consider Pauli strings $\bar a$ with the left end at a position $< x$. As in Subsec.\ \ref{sec:3e}, we assume that $\bar \rho_{\bar a}$ takes the maximally-random form at $s$ if $s$ is at a distance larger than a cut-off distance $n \gtrsim n_{\rm DW}$ to the left of the right end of the binary Pauli string $\bar a$. For such $\bar a$, the summands in Eq.\ (\ref{eq:Cxyt1}) cancel (compare with Eq.\ (\ref{eq:rhonn})). Hence, it remains to sum over binary Pauli strings $\bar a$ for which $s$ is within a distance $n$ to the left of the right end of $\bar a$ or to the right of the right end of $\bar a$. This sum can be expressed with the help of the right-propagating $n$-point densities $\bar \rho^{(n)}_{\rm }(\Delta x;t;\bar a_n,\ldots,\bar a_1)$ defined in Sec.\ \ref{sec:3e}
\begin{align}
  \label{eq:Cst}
  \langle C(s,t) \rangle =&\, 1 - \sum_{s' < s} \bar \rho^{(0)}_{\rm }(s' - t;t)
  \nonumber \\ &\, \mbox{}
  + \frac{1 - \delta(s,t)}{q^2-1}
  \bar \rho^{(0)}_{\rm }(s-t;t),
\end{align}
with
\begin{align}
  \label{eq:delta}
  \delta(s,t) =&\, 
  \sum_{m=1}^{n}
  \sum_{\bar a_{m-1},\ldots,\bar a_{1}}
  \frac{q^2-1}{\bar \rho^{(0)}_{\rm }(s-t;t)}
  \nonumber \\ &\, \mbox{} \times
  \left[ \frac{1}{q^2-1}
  \bar \rho^{(m)}_{\rm }(s-t+m;t;\XX,\bar a_{m-1},\ldots,\bar a_{1})
  \right. \nonumber \\ &\, \left. \ \ \ \ \vphantom{\frac{M}{M}}
  \mbox{} -
  \bar \rho^{(m)}_{\rm }(s-t+m;t;\II,\bar a_{m-1},\ldots,\bar a_{1})
  \right].
\end{align}
This expression for $\langle C(s,t) \rangle$, but without the term proportional to $\delta(s,t)$, was previously obtained in Ref.\ \cite{von_keyserlingk_operator_2018}. The correction term proportional to $\delta$ represents the finite width of the domain wall between the maximum-entropy and trivial phases in random circuits with non-Haar-distributed two-qubit gate operators.

In Sec.\ \ref{sec:3} we have shown that in the long-time limit, the $n$-point right-moving densities satisfy a drift-diffusion equation with drift velocity $v_{\rm B}^{(n)}$ and diffusion constant ${\cal D}^{(n)}$. We thus conclude that the same drift-diffusion process also describes the long-time behavior of the out-of-time-order-correlator (\ref{eq:OTOC}). Hence, the two terms in the first line of Eq.\ (\ref{eq:Cst}) may well be approximated as $(1/2)\mathrm{Erfc}[(s-v^{(n)}_Bt)/\sqrt{2D^{(n)}t}]$ in the long time limit, see Ref. \cite{von_keyserlingk_operator_2018}.

\begin{figure}
    \centering
\includegraphics[width=1\linewidth]{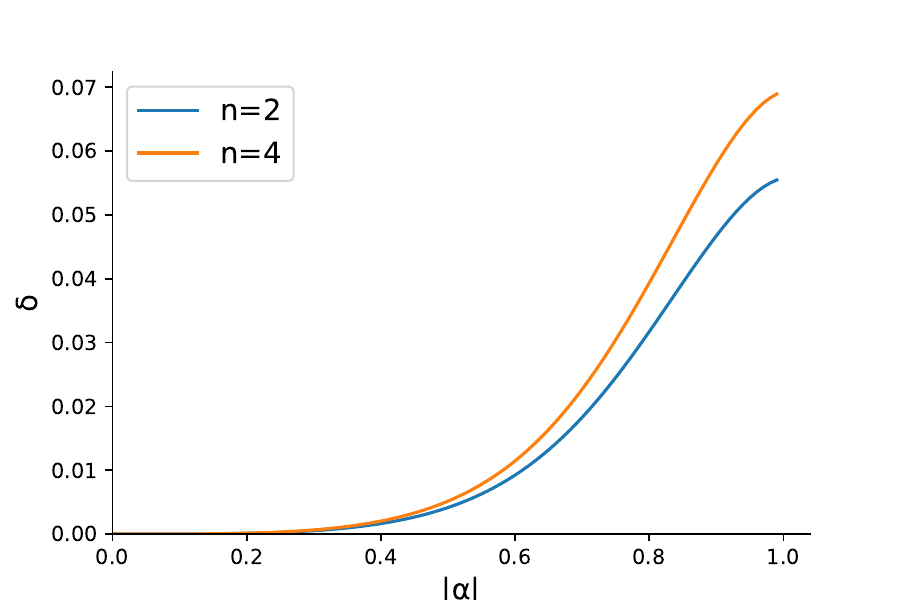}
    \caption{\justifying \small Long-time limit $\delta \equiv \lim_{t \to \infty} \delta(s,t)$ of the correction factor $\delta(s,t)$ of the out-of-time-ordered correlator for a random unitary circuit with Poisson-kernel-distributed two-qudit gates with $q=2$. The horizontal axis shows the parameter $\alpha$ of the Poisson-kernel distribution. The order $n$ of the approximation corresponds to the truncation distance, beyond which it assumed that the right-moving densities take their maximum-entropy form. }
    \label{fig:correction}
\end{figure}

\begin{figure}
    \centering
    \includegraphics[width=0.9\linewidth]{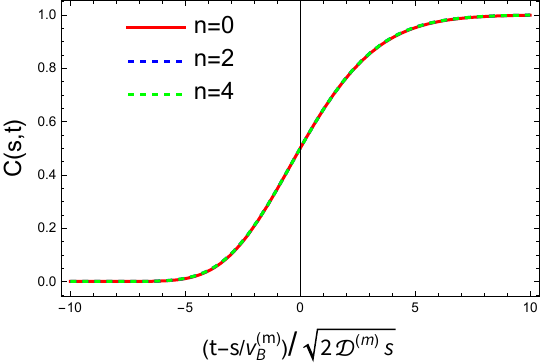}
    \caption{\justifying \small
    Out-of-time-ordered correlator $C(s,t)$ for $s=60$, $q=2$, and Poisson-kernel-distributed two-qudit gates with $\alpha=0.6$, as a function of time $t$. The three curves correspond to truncation distances $n=0$, $2$, and $4$, as indicated. }
    \label{fig:OTOC}
\end{figure}

To estimate the correction $\delta(s,t)$ in the long-time limit, we make use of the expression (\ref{eq:Rntlong}) for the large-$t$ limit of the right-moving $m$-point density $\rho^{(m)}(\Delta x;t;\bar a_m,\ldots,\bar a_1)$. Accordingly, $\rho^{(m)}(s-t+m;t;\bar a_m,\ldots,\bar a_1)$ factorizes into a function $R_{{\rm R}1}^{(m)}(s-t+m,t)$, which satisfies the drift-diffusion equation (\ref{eq:driftdiffusion}), and a function $V_1^{(n)}(\bar a_1,\ldots,\bar a_m)$, which is independent of $s$ and $t$. (The dependence on $\bar a_1,\ldots,\bar a_m$ is implicit in Eq.\ (\ref{eq:Rntlong}).) Since the width of the diffusing front is proportional to $\sqrt{t}$, it exceeds the domain-wall width $n_{\rm DW}$ for sufficiently large time $t$. Since $m \lesssim n_{\rm DW}$, this implies that $R_{{\rm R}1}^{(m)}(s-t+m,t) \to R_{{\rm R}1}^{(m)}(s-t,t)$ in the large-$t$ limit. The factor $R_{{\rm R}1}^{(n)}(s-t,t)$ then drops out from Eq.\ (\ref{eq:delta}) and we find that the large-$t$ limit of $\delta(s,t)$ is independent of both $s$ and $t$. Figure \ref{fig:correction} shows this long-time limit $\delta$ for a random unitary circuit with Poisson-kernel-distributed two-qudit gates as a function of the parameter $\alpha$. The figure shows that $\delta$ increases with $|\alpha|$, but remains numerically small even for $|\alpha| \to 1$.  As already demonstrated in Ref.\ \cite{von_keyserlingk_operator_2018}, $\bar \rho^{(0)}_{\rm }(s-t;t)$ is relatively small compared to the two terms on the first line of Eq. \eqref{eq:Cst} in the long time limit, so that the entire second-line of
Eq. \eqref{eq:Cst} is a sub-leading correction for
large $t$. Therefore, the two terms on the first line are an excellent approximation for the OTOC in the long-time limit. In Fig.\ \ref{fig:OTOC} we show the full OTOC in the long-time limit.

\section{Conclusion}
\label{sec:concl}

In this article, we have extended the study of operator spreading in random unitary circuits from circuits with Haar-distributed two-qudit gates to circuits with the more general class of unitary-invariant distributions. As in the case of a circuit with Haar-distributed gate operators, we demonstrate that the operator spreading process can be mapped to a classical stochastic growth model. The classical stochastic model satisfies drift-diffusion dynamics in the long-time limit, characterized by a drift velocity (the butterfly velocity $v_{\rm B}$) and a diffusion constant ${\cal D}$. We determine $v_{\rm B}$ and ${\cal D}$ for unitary-invariant two-qudit-gate distributions and establish a relation with the out-of-time-ordered correlator (OTOC). Unlike in the Haar-distributed case, where $v_{\rm B}$ and ${\cal D}$ are functions of the qudit dimension $q$ only and $v_{\rm B} \to 1$, ${\cal D} \to 0$ for $q \to \infty$, for the more general class of models we consider here, $v_{\rm B}$ and ${\cal D}$ depend on the precise form of the statistical distribution of the two-qudit gate operators and remain below $1$ and nonzero, respectively, in the limit $q \to \infty$.
A concrete example of a unitary-invariant distribution of the two-qudit gate operator ${\cal U}$ is the Poisson kernel \cite{KRIEGER1967375,forrester2010,brouwer_generalized_1995,mello1985a}, which is the maximum-information-entropy distribution with a fixed value of the average $\langle {\cal U} \rangle = \alpha \openone$. The Poisson kernel interpolates between the Haar-distributed circuit for $\alpha = 0$ and the trivial circuit for $\alpha = 1$.

Our solution of the stochastic model relies on the assumption that the Pauli string weights $\rho_p(t)$ approach the maximally-random form characteristic of the uniform bulk within a finite distance $n_{\rm DW}$ from the end of a Pauli string. Our explicit results for the Poisson-kernel distributed circuit supports this assumption, showing convergence assuming the maximum-entropy form for a distance $n \sim 4$ from the end of the Pauli string for all values of the parameter $\alpha$. Numerical simulations of the classical stochastic growth model, which do not require the assumption of a maximally random distribution sufficiently far from the end of the Pauli string, are in excellent agreement with our analytical calculations.

Our analysis of the classical stochastic growth model reveals three regimes of relevance to the information spreading in the random unitary circuit: At short times, the Pauli-string distribution first relaxes to a {\em binary} form, in which the Pauli-string weights $\rho_p(t)$ depend only on whether the generalized Pauli operator $\sigma_{p_x}$ at site $x$ is trivial or nontrivial. For the Haar distribution, the uniform-binary distribution is reached already after a single time step. For Poisson-kernel-distributed two-qudit gates, the characteristic time scale for the approach to the uniform-binary distribution $\tau_{\rm b}$ is typically of the order of a few time steps, although it diverges in the limit $|\alpha| \to 1$ of a trivial circuit without information spreading. In the intermediate {\em thermalization} phase, a Pauli string distribution that is localized initially spreads out in space. In this process, a domain wall separates a maximum-entropy region, in which all generalized Pauli operators have the same probability, and a trivial region, moves with the butterfly velocity $v_{\rm B}$. The time scale governing the transition from a trivial distribution to the maximally random distribution at a fixed position $x$ is $n_{\rm DW}/v_{\rm B}$, where $n_{\rm DW}$ is the width of the domain wall between trivial and maximally random regions. The fast convergence of our truncation procedure for the Poisson-kernel-distributed circuit indicates that $n_{\rm DW}$ typically is of the order of a few lattice spacings. Finally, at long time scales $\sim L/v_{\rm B}$, $L$ being the system size, an initially localized Pauli-string distribution has spread over the entire system size, marking the completion of the operator spreading process \cite{xu_scrambling_2022}. 

Although the distributions of the two-qudit gates considered here are no longer uniform in the unitary group, no symmetries were imposed on the distribution. An investigation of the role of additional symmetries on information spreading in random circuits is left for future work. 

\acknowledgments

We thank Adam Nahum, Guoyi Zhu, Vatasl Dwivedi, and Adam Chaou for their valuable discussions. Special thanks to Yiping Deng for suggestions on figure editing. Financial support was provided by the Einstein Stiftung
Berlin (Einstein Research Unit on Quantum Devices).

\appendix
\section{Generalized Pauli matrices}
\label{app:1}

The generalized Pauli matrices $\sigma_a$, $a =0,1,\ldots,q^2-1$, form a basis for operators acting on a qudit, a quantum system with $q$ degrees of freedom $|m\rangle$, $m=0,1,\ldots,q-1$. We represent the index $a = (a',a'')$ as a pair of two integers $a'$, $a'' \in \mathbb{Z}_q$. The complete family of $q^2$ independent generalized Pauli matrices is then defined as \cite{patera_pauli_1988}
\begin{equation}
\sigma_{a}=\sum_{m=0}^{q-1}\omega^{ma''}\left|m+a'\right>\left<m\right|,
\end{equation}
where $\omega=\mathrm{e}^{\frac{2\pi i}{q}}$ and $m + a'$ is taken $\mod q$.
The generalized Pauli matrices satisfy the orthonormality relation
\begin{align}
  \tr\sigma_{a}\sigma_{b}^\dagger= q \delta_{a,b}
\end{align}
and the commutation relation
\begin{align}
  \sigma_{a}\sigma_{b}=\omega^{b'a''-a'b''}\sigma_{b}\sigma_{a}.
\end{align}
They also satisfy the completeness relation
\begin{equation}
  \sum_{a} \mbox{tr}\, A \sigma^\dagger_{a}\, \mbox{tr}\, B^\dagger\sigma_{a}= q \mbox{tr}\, A B^\dagger  \label{eq:complete}
\end{equation}
for arbitrary $q \times q$ matrices $A$ and $B$.
Because of Eq.\ (\ref{eq:complete}), the generalized Pauli matrices $\sigma_{a}$ are also said to be a unitary $1$-design \cite{roberts_chaos_2017}. 

Since the evolution operators act on pairs of qudits, we will often consider the tensor product $\sigma_{a_x} \otimes \sigma_{a_y}$ of generalized Pauli matrices for two qudits at positions $x$ and $y$. With the shorthand notations $a = (a_x,a_y) = (a_x',a_x'',a_y',a_y'')$ and $\Sigma_{a} = \sigma_{a_x} \otimes \sigma_{a_y}$, we observe that the $q^4$ matrices $\Sigma_{a}$ satisfy the orthonormality relation
\begin{equation}
  \tr \Sigma_a \Sigma_b^{\dagger} = q^2 \delta_{a,b},
  \label{eq:SigmaOrthogonal}
\end{equation}
whereas their commutation relation is
\begin{equation}
  \Sigma_a \Sigma_b = - \varphi(a,b) \Sigma_b \Sigma_a,
  \label{eq:SigmaCommutation}
\end{equation}
with 
\begin{equation}
  \varphi(a,b) = - \omega^{b_x'a_x''-a_x'b_x'' + b_y'a_y''-a_y'b_y''}.
  \label{eq:phidefinition}
\end{equation}
The function $\varphi(a,b)$ satisfies
\begin{equation}
  \varphi(0,0,b_x,b_y) = -1
  \label{eq:phispecial}
\end{equation}
and
\begin{equation}
\label{eq:phihermitian}
  \varphi(a,b) = \varphi(b,a)^*,
\end{equation}
as well as the sum rules
\begin{align}
  \label{eq:phiidentities}
  \sum_{a_x \neq 0} \varphi(a_x,0,b_x,b_y) =&\,
  1 - q^2 \delta_{b_x,0}, \nonumber \\
  \sum_{a_y \neq 0} \varphi(0,a_y,b_x,b_y) =&\,
  1 - q^2 \delta_{b_y,0}, \\
  \nonumber
  \sum_{a_x \neq 0} \sum_{a_y \neq 0} \varphi(a_x,a_y,b_x,b_y) =&\,
  - 1 + q^2 (\delta_{b_x,0} + \delta_{b_y,0})
  \nonumber \\ &\, \mbox{} 
  - q^4 \delta_{b_x,0} \delta_{b_y,0}.
  \nonumber
\end{align}
Because of the hermiticity relation (\ref{eq:phihermitian}) analogous sum rules also apply to summations of $\varphi(a_x,a_y,b_x,b_y)$ over the second index pair $b_x$, $b_y$.

\section{Calculation of $\langle {\cal W}_{ap;x,y} {\cal W}_{bp;x,y}^* \rangle$}
\label{app:2}
\begin{figure*}[t!]
\centering
\begin{tikzpicture}[every node/.style={draw, circle, minimum size=4.5pt}, every label/.style=draw,node distance=6mm]
    \node[fill, inner sep=0pt, blue](n1) at (0,0) {};
    \node[inner sep=0pt,below=of n1,blue] (n2)  {};
   \node[below=of n2, inner sep=0pt, blue](n3)  {};
    \node[fill, inner sep=0pt,below=of n3,blue] (n4)   {};
     \node[fill, inner sep=0pt, blue](n5) at (1,0) {};
    \node[inner sep=0pt,below=of n5,blue] (n6)  {};
     \node[ inner sep=0pt,below=of n6, blue](n7)  {};
    \node[fill,inner sep=0pt,below=of n7,blue] (n8) {};
    \node[rectangle](n9) at (0.5, -2.8){};
    \node[rectangle](n10) at (0.5, 0.5){};
    \draw[double distance between line centers=0.9mm,  blue, line width=.2mm] (n1) to  (n2);
    \draw [middlearrow={>[length=1.4mm,line
     width=1pt]}, black, line width=.5mm]  (n2) to (n3);
     \draw[double distance between line centers=0.9mm,  blue, line width=.2mm] (n3) to node [xshift=-0.2cm,draw=none]  {*}(n4);
     \draw [double distance between line centers=0.9mm,  blue, line width=.2mm] (n7) to (n8);
     \draw [double distance between line centers=0.9mm,  blue, line width=.2mm] (n5) to node [xshift=0.2cm, draw=none]{*} (n6);
    \draw [middlearrow={<[length=1.4mm,line
     width=1pt]}, black, line width=.5mm]  (n6) to (n7);
     \draw [line width=0.3mm] (n1) to [bend left] (n10.west); 
     \draw [line width=0.3mm] (n10.east) to [bend left] (n5); 
     \draw [line width=0.3mm] (n4) to [bend right] (n9.west); 
     \draw [line width=0.3mm] (n9.east) to [bend right] (n8); 
      \draw [densely dashed, black, line width=.5mm]  (n1) to (n5);
      \draw [densely dashed, black, line width=.5mm]  (n2) to (n6);
      \draw [densely dashed, black, line width=.5mm]  (n3) to (n7);
      \draw [densely dashed, black, line width=.5mm]  (n4) to (n8);
      \node[fill, inner sep=0pt, blue](n1) at (2,0) {};
    \node[inner sep=0pt,below=of n1,blue] (n2)  {};
   \node[below=of n2, inner sep=0pt, blue](n3)  {};
    \node[fill,inner sep=0pt,below=of n3,blue] (n4) {};
     \node[fill, inner sep=0pt, blue](n5) at (3,0) {};
    \node[inner sep=0pt,below=of n5,blue] (n6)  {};
     \node[inner sep=0pt,below=of n6, blue](n7) {};
    \node[fill,inner sep=0pt,below=of n7,blue] (n8) {};
    \node[rectangle](n9) at (2.5, -2.8){};
    \node[rectangle](n10) at (2.5, 0.5){};
    \draw[double distance between line centers=0.9mm,  blue, line width=.2mm] (n1) to  (n2);
    \draw [middlearrow={>[length=1.4mm,line
     width=1pt]}, black, line width=.5mm]  (n2) to (n3);
     \draw[double distance between line centers=0.9mm,  blue, line width=.2mm] (n3) to node [xshift=-0.2cm,draw=none]  {*}(n4);
     \draw [double distance between line centers=0.9mm,  blue, line width=.2mm] (n7) to (n8);
     \draw [double distance between line centers=0.9mm,  blue, line width=.2mm] (n5) to node [xshift=0.2cm, draw=none]{*} (n6);
    \draw [middlearrow={<[length=1.4mm,line
     width=1pt]}, black, line width=.5mm]  (n6) to (n7);
     \draw [line width=0.3mm] (n1) to [bend left] (n10.west); 
     \draw [line width=0.3mm] (n10.east) to [bend left] (n5); 
     \draw [line width=0.3mm] (n4) to [bend right] (n9.west); 
     \draw [line width=0.3mm] (n9.east) to [bend right] (n8); 
     \draw [densely dashed, black, line width=.5mm]  (n1) to (n5);
      \draw [densely dashed, black, line width=.5mm]  (n2) to (n6);
      \draw [densely dashed, black, line width=.5mm]  (n3) to (n7);
      \draw [densely dashed, black, line width=.5mm]  (n4) to (n8);
      \node[draw=none](n1) at (3.5,-1.3) {};
    \node [draw=none](n2) at (5,-1.3) {};
     \draw[double distance between line centers=0.9mm, , black, line width=.2mm, -{>[length=1.5mm, width=3mm]}] (n1) to  (n2);
      \node[fill, inner sep=0pt, blue](n1) at (5.5,0) {};
    \node[inner sep=0pt,below=of n1,blue] (n2)  {};
   \node[below=of n2, inner sep=0pt, blue](n3)  {};
    \node[fill, inner sep=0pt,below=of n3,blue] (n4)   {};
     \node[fill, inner sep=0pt, blue](n5) at (6.5,0) {};
    \node[inner sep=0pt,below=of n5,blue] (n6)  {};
     \node[ inner sep=0pt,below=of n6, blue](n7)  {};
    \node[fill,inner sep=0pt,below=of n7,blue] (n8) {};
    \draw[double distance between line centers=0.9mm,  blue, line width=.2mm] (n1) to  (n2);
     \draw[double distance between line centers=0.9mm,  blue, line width=.2mm] (n3) to node [xshift=-0.2cm,draw=none]  {*}(n4);
     \draw [double distance between line centers=0.9mm,  blue, line width=.2mm] (n7) to (n8);
     \draw [double distance between line centers=0.9mm,  blue, line width=.2mm] (n5) to node [xshift=0.2cm, draw=none]{*} (n6);
      \draw [densely dashed, black, line width=.5mm]  (n1) to node [xshift=1cm,yshift=1cm, draw=none]  {type-i subdiagram}(n5);
      \draw [densely dashed, black, line width=.5mm]  (n2) to (n6);
      \draw [densely dashed, black, line width=.5mm]  (n3) to (n7);
      \draw [densely dashed, black, line width=.5mm]  (n4) to (n8);
      \node[fill, inner sep=0pt, blue](n1) at (7.5,0) {};
    \node[inner sep=0pt,below=of n1,blue] (n2)  {};
   \node[below=of n2, inner sep=0pt, blue](n3)  {};
    \node[fill, inner sep=0pt,below=of n3,blue] (n4)   {};
     \node[fill, inner sep=0pt, blue](n5) at (8.5,0) {};
    \node[inner sep=0pt,below=of n5,blue] (n6)  {};
     \node[ inner sep=0pt,below=of n6, blue](n7)  {};
    \node[fill,inner sep=0pt,below=of n7,blue] (n8) {};
    \draw[double distance between line centers=0.9mm,  blue, line width=.2mm] (n1) to  (n2);
     \draw[double distance between line centers=0.9mm,  blue, line width=.2mm] (n3) to node [xshift=-0.2cm,draw=none]  {*}(n4);
     \draw [double distance between line centers=0.9mm,  blue, line width=.2mm] (n7) to (n8);
     \draw [double distance between line centers=0.9mm,  blue, line width=.2mm] (n5) to node [xshift=0.2cm, draw=none]{*} (n6);
      \draw [densely dashed, black, line width=.5mm]  (n1) to (n5);
      \draw [densely dashed, black, line width=.5mm]  (n2) to (n6);
      \draw [densely dashed, black, line width=.5mm]  (n3) to (n7);
      \draw [densely dashed, black, line width=.5mm]  (n4) to (n8);
      \node[fill, inner sep=0pt, blue](n1) at (10,0) {};
    \node[inner sep=0pt,below=of n1,blue] (n2)  {};
   \node[below=of n2, inner sep=0pt, blue](n3)  {};
    \node[fill,inner sep=0pt,below=of n3,blue] (n4) {};
     \node[fill, inner sep=0pt, blue](n5) at (11,0) {};
    \node[inner sep=0pt,below=of n5,blue] (n6)  {};
     \node[inner sep=0pt,below=of n6, blue](n7) {};
    \node[fill,inner sep=0pt,below=of n7,blue] (n8) {};
    \node[rectangle](n9) at (10.5, -2.8){};
    \node[rectangle](n10) at (10.5, 0.5){};
    \draw [middlearrow={>[length=1.4mm,line
     width=1pt]}, black, line width=.5mm]  (n2) to (n3);
    \draw [middlearrow={<[length=1.4mm,line
     width=1pt]}, black, line width=.5mm]  (n6) to (n7);
     \draw [line width=0.3mm] (n1) to [bend left] (n10.west); 
     \draw [line width=0.3mm] (n10.east) to [bend left] (n5); 
     \draw [line width=0.3mm] (n4) to [bend right] (n9.west); 
     \draw [line width=0.3mm] (n9.east) to [bend right] (n8); 
     \draw [densely dashed, black, line width=.5mm]  (n1) to node [xshift=1cm,yshift=1cm, draw=none]  {type-ii subdiagram} (n5);
      \draw [densely dashed, black, line width=.5mm]  (n2) to (n6);
      \draw [densely dashed, black, line width=.5mm]  (n3) to (n7);
      \draw [densely dashed, black, line width=.5mm]  (n4) to (n8);
       \node[fill, inner sep=0pt, blue](n1) at (12,0) {};
    \node[inner sep=0pt,below=of n1,blue] (n2)  {};
   \node[below=of n2, inner sep=0pt, blue](n3)  {};
    \node[fill,inner sep=0pt,below=of n3,blue] (n4) {};
     \node[fill, inner sep=0pt, blue](n5) at (13,0) {};
    \node[inner sep=0pt,below=of n5,blue] (n6)  {};
     \node[inner sep=0pt,below=of n6, blue](n7) {};
    \node[fill,inner sep=0pt,below=of n7,blue] (n8) {};
    \node[rectangle](n9) at (12.5, -2.8){};
    \node[rectangle](n10) at (12.5, 0.5){};only
    \draw [middlearrow={>[length=1.4mm,line
     width=1pt]}, black, line width=.5mm]  (n2) to (n3);
    \draw [middlearrow={<[length=1.4mm,line
     width=1pt]}, black, line width=.5mm]  (n6) to (n7);
     \draw [line width=0.3mm] (n1) to [bend left] (n10.west); 
     \draw [line width=0.3mm] (n10.east) to [bend left] (n5); 
     \draw [line width=0.3mm] (n4) to [bend right] (n9.west); 
     \draw [line width=0.3mm] (n9.east) to [bend right] (n8); 
     \draw [densely dashed, black, line width=.5mm]  (n1) to (n5);
      \draw [densely dashed, black, line width=.5mm]  (n2) to (n6);
      \draw [densely dashed, black, line width=.5mm]  (n3) to (n7);
      \draw [densely dashed, black, line width=.5mm]  (n4) to (n8);
\end{tikzpicture}
\caption{\justifying \small Left: Example of a diagram contributing to the average $\langle {\cal W}_{ap;x,y} {\cal W}_{bq;x,y} \rangle$. Right: Type-i subdiagram with dashed lines and double lines only and type-ii subdiagram with dashed and solid lines only. The contribution to $\langle {\cal W}_{ap;x,y} {\cal W}_{bq;x,y} \rangle$ is found by inspecting the closed loops in each subdiagram.
}
\label{example}
\vspace{0cm}
\end{figure*}
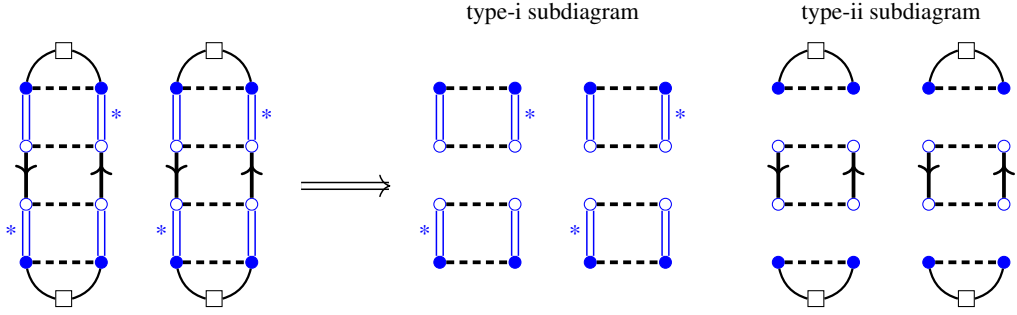

The ensemble average $\langle W_{ap}W^{*}_{bq} \rangle$ can be expressed as a product of pairwise averages,
\begin{equation}
  \left<W_{ap}W^{*}_{bq}\right>=\prod_{\text{x even}}
  \left\{\begin{array}{l}
  \left<\mathcal{W}_{a p; x,x+1} \mathcal{W}^*_{b q; x,x+1}\right>\text { t even, } \\
  \left<\mathcal{W}_{a p; x-1, x }\mathcal{W}^*_{b q; x-1, x } \right> \text { t odd, }
\end{array}\right.
\end{equation}
The matrices ${\cal W}_{ap;x,x+1}$ can be expressed in terms of the pair-qudit evolution matrix ${\cal U}_{x,x+1}$, see Eq.\ (\ref{eq:WU}),
\textcolor{black}{
\begin{align}
  \label{eq:WUapp}
  {\cal W}_{ap;x,y} =&\,
  \frac{1}{q^2}
  \mbox{tr}\, {\cal U}^{\dagger}_{x,y}
  \Sigma_a
  {\cal U}_{x,y}
  \Sigma_p^{\dagger},
\end{align}}
where $\Sigma_a = \sigma_{a_x} \otimes \sigma_{a_y}$ and $\Sigma_p = \sigma_{p_x} \otimes \sigma_{p_y}$, see App.\ \ref{app:1}.
For a pair evolution matrix ${\cal U}_{x,y}$ that has a distribution that is invariant under unitary transformations, we may replace ${\cal U}_{x,y}$ by
\begin{equation}
  {\cal U}_{x,y} \to \mathcal{V}_{x, y} {\cal U}_{x,y} \mathcal{V}^{\dagger}_{x, y},
  \label{eq:Usubst}
\end{equation}
with ${\cal V}_{x,x+1}$ a unitary matrix that can be chosen arbitrarily. In this appendix, we first present a simple argument in support of Eq.\ (\ref{eq:WW}), making a specific choice for the matrices ${\cal V}_{x,x+1}$. After that, we take the matrices ${\cal V}_{x,x+1}$ to be Haar-distributed random matrices that are statistically independent of ${\cal U}_{x,x+1}$ and perform an average over ${\cal V}_{x,x+1}$. This allows us to derive Eq.\ \ref{eq:Wgeneral}), which expresses the ensemble average $\langle {\cal W}_{ap} {\cal W}_{bp}^* \rangle$ in terms of the four moments of the pair-qudit evolution operator ${\cal U}_{x,x+1}$ defined in Eq.\ (\ref{eq:R112}) of the main text.

\textcolor{black}{
{\em Proof of Eq.\ (\ref{eq:WW}).---} After the substitution (\ref{eq:Usubst}), we can rewrite the average $\langle {\cal W}_{ap;x,y} {\cal W}_{bp;x,y}^* \rangle$ as
\begin{align}
  \langle {\cal W}_{ap;x,y} {\cal W}_{bp;x,y}^* \rangle &=
  \frac{1}{q^4}
  \mbox{tr}\, {\cal U}^{\dagger}_{x,y}
  \widetilde\Sigma_a
  {\cal U}_{x,y}
  \widetilde\Sigma_p^{\dagger}
  \mbox{tr}\, {\cal U}^{\dagger}_{x,y}
  \widetilde\Sigma^\dagger_b
  {\cal U}_{x,y}
  \widetilde\Sigma_p
\end{align}
where 
\begin{equation}
  \widetilde \Sigma_a=\mathcal{V}_{x, y} \Sigma_a \mathcal{V}^{\dagger}_{x, y}.
\end{equation}
We make the special choice that $\mathcal{V}_{x, y} = \Sigma_q$ is a generalized Pauli matrix. From Eq.\ (\ref{eq:SigmaCommutation}), we then find, that
\begin{equation}
  \widetilde\Sigma_a=-\varphi^*(a,q)\Sigma_a,
\end{equation}
with $\varphi(a,q)$ defined in Eq.\ (\ref{eq:phidefinition}). It follows that
\begin{equation}
  \langle {\cal W}_{ap;x,y} {\cal W}_{bp;x,y}^* \rangle=
  -\varphi(b-a,q)\langle {\cal W}_{ap} {\cal W}_{bp}^* \rangle,
\end{equation}
where we made use of the property $\varphi(a,q)^* \varphi(b,q) = -\varphi(b-a,q)$, which immediately follows from the definition (\ref{eq:phidefinition}). Because $\Sigma_q$ can be chosen arbitrary, we conclude that $\langle {\cal W}_{ap;x,y} {\cal W}_{bp;x,y}^* \rangle$ can be nonzero only if $\Sigma_a=\Sigma_b$, which immediately leads to Eq. (\ref{eq:WW}).}

{\em Proof of Eq.\ (\ref{eq:Wgeneral}).---}
To derive Eq.\ (\ref{eq:Wgeneral}), we use the
diagrammatic method developed by Beenakker and one of the authors \cite{brouwer_diagrammatic_1996} to perform the average over $\equiv {\cal V}_{x,x+1}$. Following Ref.\ \cite{brouwer_diagrammatic_1996}, the elements of the Haar-distributed matrix ${\cal V}_{x,x+1}$ are represented by double lines,
\begin{equation}
\begin{tikzpicture}[every node/.style={draw, circle, minimum size=4.5pt},node distance=6mm]
    \node[fill, inner sep=0pt, blue](n1) at (0,0.1) {};
    \node[inner sep=0pt,below=of n1,blue] (n2) at (0,0) {};
     \draw[double distance between line centers=0.9mm, middlearrow={>[length=1.4mm,line width=1pt,open]}, blue, line width=.2mm] (n1) to node [left, draw=none]{${\cal V}=$} (n2);
     
    \node[fill, inner sep=0pt, blue](n3) at (2,0.1) {};
    \node[inner sep=0pt,below=of n1,blue] (n4) at (2,0) {};
    \draw[double distance between line centers=0.9mm, middlearrow={<[length=1.4mm,line width=1pt,open]}, blue, line width=.2mm]  (n3) to node [left, draw=none]{${\cal V}^*=$} node [auto, draw=none]{*.} (n4);
\end{tikzpicture}
\end{equation}
(We suppressed the spatial indices $x$ and $x+1$ to keep the notation simple.)
Elements of the evolution matrix ${\cal U}_{x,x+1}$, as they appear on the r.h.s.\ of the substitution (\ref{eq:Usubst}), are represented as single lines with arrows,
\begin{equation}
\begin{tikzpicture}[node distance=8mm]
    \node(n1) at (0,0.1) {};
   \node[below=of n1](n2) at (0,0) {};
    \draw [middlearrow={>[length=1.4mm,line
     width=1pt]}, black, line width=.5mm]  (n1) to node [left, draw=none]{${\cal U}=\ $} (n2);
     \node(n3) at (2,0.1) {};
    \node[below=of n1] (n4) at (2,0) {};
      \draw[ middlearrow={<[length=1.4mm,line
     width=1pt]}, black, line width=.5mm]  (n3) to node [left, draw=none]{${\cal U}^*=\ $} node [auto, draw=none]{*} (n4);
\end{tikzpicture},
\end{equation}
whereas the generalized Pauli matrices $\Sigma_a$, $\Sigma_b$, $\Sigma_p$, and $\Sigma_q$ are represented by solid lines with an open square. Only terms in which the first and second indices of all factors ${\cal V}$ and ${\cal V}^*$ are pairwise equal contribute to $\langle {\cal W}_{ap;x,x+1} {\cal W}_{bq;x,x+1} \rangle$. Diagrammatically, this is indicated by contractions, represented by dashed lines, between pairs of filled dots (for the first indices) and pairs of open dots (for the second indices). An example of a diagram contributing to $\langle {\cal W}_{ap;x,x+1} {\cal W}_{bq;x,x+1} \rangle$ is shown in Fig.\ \ref{example}.
To evaluate the contribution to the average for each contraction, one considers two subdiagrams: (i) the diagram with dashed and dotted lines only, and (ii) the diagram with dashed and solid lines only, see Fig.\ \ref{example}. Subdiagram (i) determines the Weingarten number \cite{weingarten1978,collins2003}. It consists of closed loops formed out of alternating dashed lines (for the contractions) and double lines (for the Haar matrices $V$ and $V^*$). Defining the ``length'' $c_i$ of such a loop as the number of double lines representing ${\cal V}$, where $i=1,\ldots,k$, $k$ being the number of such loops, the Weingarten number of the diagram is $V_{c_1,\ldots,c_k}$. The five Weingarten numbers that appear in our calculation are \cite{brouwer_diagrammatic_1996,collins2003}
\begin{align}
  \label{eq:VWeingarten}
  V_{1,1,2} =&\, \frac{4 q^2 - q^6}{q^4(q^4-1)(q^4-4)(q^4-9)}, \nonumber \\
  V_{2,2} =&\, \frac{q^4 + 6}{q^4(q^4-1)(q^4-4)(q^4-9)}, \nonumber \\
  V_{1,3} =&\, \frac{2 q^4 - 3}{q^4(q^4-1)(q^4-4)(q^4-9)}, \nonumber \\
  V_{1,1,1,1} =&\, \frac{q^8 - 8 q^4 + 6}{q^4(q^4-1)(q^4-4)(q^4-9)}, \nonumber \\
  V_{4} =&\, -\frac{5 q^2}{q^4(q^4-1)(q^4-4)(q^4-9)}.
\end{align}
Likewise, subdiagram (ii) contains loops consisting of alternating single lines with arrows (for ${\cal U}$ and ${\cal U}^*$) and dashed lines connecting open dots and loops consisting of alternating single lines with squares (for the generalized Pauli matrices $\Sigma_a$, $\Sigma_b$, $\Sigma_p$, and $\Sigma_q$) and dashed lines connecting open dots. Each of these loops contributes the trace of the product of matrices ${\cal U}$ and ${\cal U}^*$ or of generalized Pauli matrices $\Sigma_a$, $\Sigma_b$, $\Sigma_p$, and $\Sigma_q$, in the order in which they appear in the loop \cite{brouwer_diagrammatic_1996}. For the example in Fig.\ \ref{example}, there are four loops of length $1$ in subdiagram (i), so that the Weingarten number is $V_{1,1,1,1}$, whereas the type-ii loops give a factor $(\tr {\cal U} {\cal U}^{\dagger})^4 \tr \Sigma_a \tr \Sigma_b \tr \Sigma_p \tr \Sigma_q$. Using $\tr {\cal U} {\cal U}^{\dagger} = q^2$, $\tr \Sigma_a = q^2 \delta_{a,0}$ and taking into account an additional prefactor $1/q^4$ from Eq.\ (\ref{eq:WUapp}), we then find that the contribution of the entire diagram is $q^{8} V_{1,1,1,1} \delta_{a,0} \delta_{b,0} \delta_{p,0} \delta_{q,0}$.\\
\begin{figure*}[tb]
\centering
\vskip -3 cm
\includegraphics[scale=1,trim=2cm 1.5cm 2cm 1cm, clip]{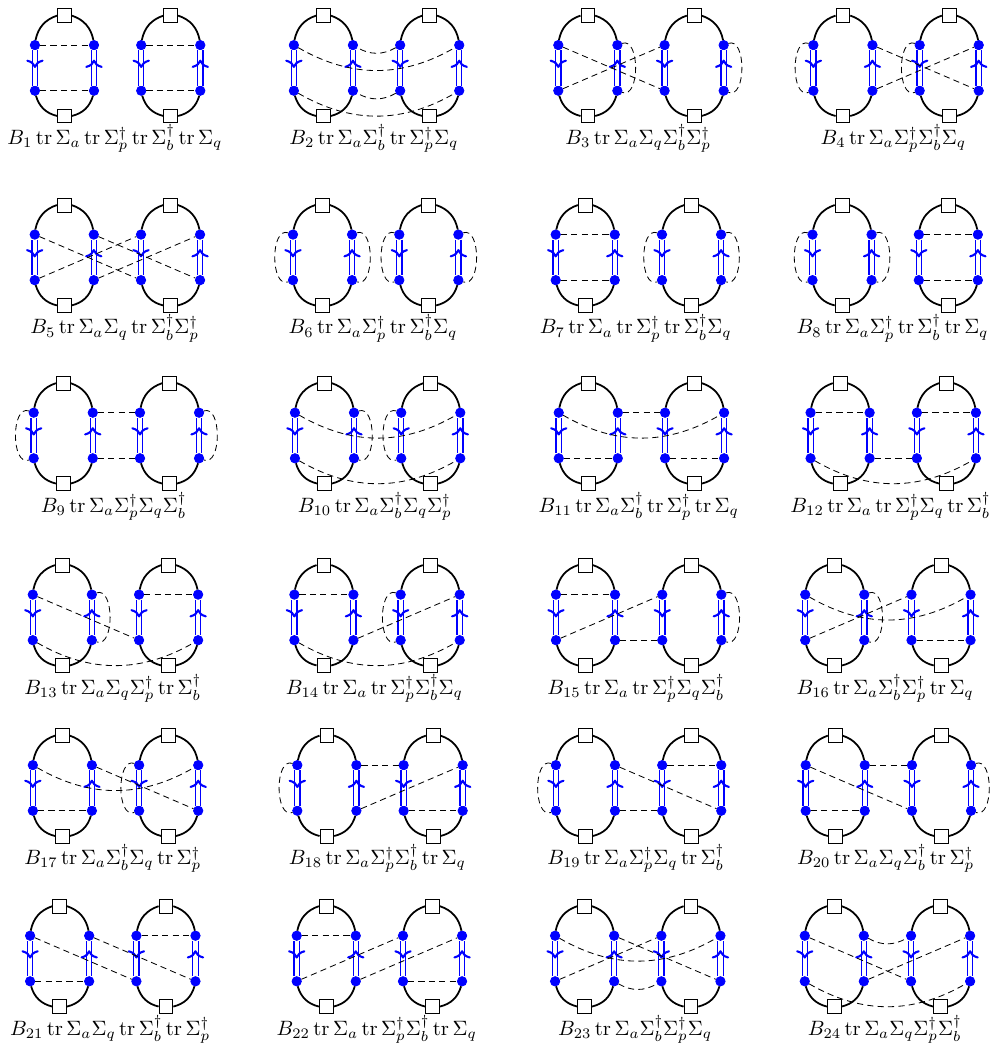}
\vskip -4cm
\caption{\justifying \small
Contractions of closed dots (corresponding to the generalized Pauli matrices) and their contribution to $\langle {\cal W}_{ap;x,y} {\cal W}_{bq;x,y} \rangle$. The factors $B_j$, $j=1,2,\ldots,24$ refer to the contributions from contractions of open dots, each of which involves a sum over $24$ contributions. The first diagram with contribution $B_{1} \tr \Sigma_{a} \tr \Sigma_{p}^{\dagger} \tr \Sigma_{b}^{\dagger} \tr \Sigma_{q}$ corresponds to that of Fig.\ \ref{example}.}
\label{fig:diagrams}
\end{figure*}

There are 24 possible contractions of open dots and 24 possible contractions of closed dots, yielding a total of 576 diagrams in the diagrammatic evaluation of $\langle {\cal W}_{ap;x,y} {\cal W}_{bq;x,y} \rangle$. We organize these diagrams according to the contractions of the closed dots, which determines the appearance of Kronecker deltas involving the indices $a$, $b$, $p$, and $q$. The 24 possibilities and their contributions to $\langle {\cal W}_{ap;x,y} {\cal W}_{bq;x,y} \rangle$ are shown in Fig.\ \ref{fig:diagrams}. Each contribution contains a factor $B_j$, $j=1,2,\ldots,24$, which represents the 24 possible contractions of the open dots and a product of traces over generalized Pauli matrices $\Sigma_a$, $\Sigma_b^{\dagger}$, $\Sigma_p^{\dagger}$, and $\Sigma_q$ or products thereof. We now discuss the coefficients $B_j$ and the products of traces of generalized Pauli matrices separately.

Instead of calculating each of the coefficients $B_j$, $j=1,2,\ldots,24$, individually, which requires the evaluation of $24$ diagrams each, we proceed as follows: We first observe that many of the diagrams in Fig.\ \ref{fig:diagrams} have the same structure, so that some of the coefficients $B_j$ are identical,
\begin{align}
  \label{eq:Bconstraint1}
  & B_{1}=B_{2}, \nonumber \\
  & B_{3}=B_{4}^*, \nonumber \\
  & B_{7}=B_{8}=B_{9}=B_{10}, \nonumber \\
  & B_{11}=B_{12}, \nonumber \\
  & B_{13}=B_{14}^*=B_{15}=B_{16}=B_{17}^*=B_{18}^*=B_{19}^*=B_{20}, \nonumber \\
  & B_{21}=B_{22}=B_{23}=B_{24}.
\end{align}
Moreover, since $\langle {\cal W}_{ap;x,y} {\cal W}_{bq;x,y} \rangle = 0$ if $a = 0$ and $p \neq 0$, see Eq.\ (\ref{eq:special}), we have
\begin{align}
  0 =&\,
  (B_{2} + q^2B_{12} + B_{13} + B_{19})
  \tr \Sigma_b^{\dagger} \tr \Sigma_p^{\dagger} \Sigma_q
  \nonumber \\ &\, \mbox{} +
  (B_{5} + B_{16} + B_{18} + q^2B_{22})
  \tr \Sigma_p^{\dagger} \Sigma_b^{\dagger} \tr \Sigma_q
  \nonumber \\ &\, \mbox{} +
  (B_{3} + B_{9} + q^2B_{15} + B_{23})
   \tr \Sigma_p^{\dagger} \Sigma_q \Sigma_b^{\dagger}
  \nonumber \\ &\, \mbox{} +
  (B_{4} + B_{10} + q^2B_{14} + B_{24}) 
  \tr \Sigma_b^{\dagger} \Sigma_q \Sigma_p^{\dagger}.
  \label{eq:Bconstraint2}
\end{align}
Since this relation holds for arbitrary choice of $\Sigma_b$ and $\Sigma_q$, each of the factors between brackets must be zero individually. Finally, from $\langle {\cal W}_{ap;x,y} {\cal W}_{bq;x,y} \rangle = 1$ if $a = b = p = q = 0$, see Eq.\ (\ref{eq:special}), we conclude that
\begin{align}
  1 =&\, q^8 B_{1} + q^6 (B_{7} + B_{8} + B_{11} + B_{12} + B_{21} + B_{22})
  \nonumber \\ &\, \mbox{}
  + q^4 (B_{2} + B_{5} + B_{6} + B_{13} + B_{14} + B_{15} 
  \nonumber \\ &\, \ \ \ \ \mbox{}
  + B_{16} + B_{17} + B_{18} + B_{19} + B_{20})
  \nonumber \\ &\, \mbox{}
  + q^2 (B_{9} + B_{10} + B_{3} + B_{4} + B_{23} + B_{24}). 
  \label{eq:Bconstraint3}
\end{align}

After taking into account these constraints, only $B_{3}$, $B_{5}$, and $B_{6}$ remain as independent coefficients. These can be calculated explicitly using the diagrammatic method. The result of this calculation is
\begin{align}
  \label{eq:Bresult}
  B_{3} =&\,
  q^{-4} (\mathcal{R}_{2,2}+4\mathcal{R}_{1;1}+\mathcal{R}_{1,1;1,1}) V_{1,1,2}
  \nonumber  \\ &\, \mbox{}
  + q^{-2} (2+ q^{-2} \mathcal{R}^{*}_{1,1;2}) V_{2,2}
  + q^{-4} \mathcal{R}_{1,1;2} V_{1,1,1,1}
  \nonumber \\ &\, \mbox{}
  + 4 q^{-2} (\mathcal{R}_{1;1} + 1) V_{1,3}
  + 2 (2 q^{-4} \mathcal{R}_{1;1}+ 1) V_4 
  , 
  \end{align}
  \begin{align}
  B_{5} =&\, 
  2 q^{-2} (2 + q^{-2} \mbox{Re}\, \mathcal{R}_{1,1;2}) V_{1,1,2}
  + 8 q^{-4} \mathcal{R}_{1;1} V_{1,3}
  \nonumber \\ &\, \mbox{}
  + (2+ q^{-4} \mathcal{R}_{1,1;1,1}) V_{2,2}
  + 2 q^{-2} (1+2 \mathcal{R}_{1;1}) V_4
  \nonumber \\ &\, \mbox{}
  + q^{-4} \mathcal{R}_{2,2} V_{1,1,1,1}
  ,
  \end{align}
  \begin{align}
  B_{6} =&\, 
  2 q^{-2} (2 \mathcal{R}_{1;1} + q^{-2} \Re\, \mathcal{R}_{1,1;2} ) V_{1,1,2}
  \nonumber \\ &\, \mbox{}
  + 8 q^{-4} \mathcal{R}_{1;1} V_{1,3} 
  + (2+ q^{-4} \mathcal{R}_{2,2}) V_{2,2}
  \nonumber \\ &\, \mbox{}
  + q^{-4} \mathcal{R}_{1,1;1,1} V_{1,1,1,1}
  + 6 q^{-2} V_4,
\end{align}
where the moments ${\cal R}_{1;1}$, ${\cal R}_{2;2}$, ${\cal R}_{1,1;1,1}$, and ${\cal R}_{1,1;2}$ were defined in Eq.\ (\ref{eq:R112}) of the main text. The coefficient $B_{3}$ is complex; $B_{5}$ and $B_{6}$ are real.

For the calculation of the traces over products of up to two generalized Pauli matrices, we now specialize to the case $p = q$, which is the case of interest for the calculation in the main text. Traces containing one or two generalized Pauli matrices can be calculated with the help of the orthonormality relation (\ref{eq:SigmaOrthogonal}), which gives $\tr \Sigma_a = q^2 \delta_{a,0}$, $\tr \Sigma_a \Sigma_b^{\dagger} = q^2 \delta_{a,b}$, $\tr \Sigma_a \Sigma_p = q^2 \delta_{a,-p}$, etc. For a trace with a product of three generalized Pauli matrices, we similarly have $\tr \Sigma_a \Sigma_b^{\dagger} \Sigma_p^{\dagger} \tr \Sigma_p = q \delta_{p,0} \tr \Sigma_a \Sigma_b^{\dagger} = q^2 \delta_{p,0} \delta_{a,b}$ and $\tr \Sigma_a \Sigma_p^{\dagger} \Sigma_p \tr \Sigma_b^{\dagger} = q \delta_{b,0} \tr \Sigma_a = q^2 \delta_{a,0} \delta_{b,0}$, etc. Finally, a trace involving four generalized Pauli matrices can be evaluated using the commutation relation (\ref{eq:SigmaCommutation}), {\em e.g.}, $\tr \Sigma_a \Sigma_p \Sigma_b^{\dagger} \Sigma_p^{\dagger} = -\varphi(a,p) \tr \Sigma_p \Sigma_a \Sigma_b^{\dagger} \Sigma_p^{\dagger} = -\varphi(a,p) q^2 \delta_{a,b}$. After evaluating all traces in this manner, all terms contributing to $\langle {\cal W}_{ap;x,y} {\cal W}_{bp;x,y} \rangle$ either contain the Kronecker delta symbol $\delta_{a,b}$ or the product $\delta_{a,0} \delta_{b,0}$. In either case, $\langle {\cal W}_{ap;x,y} {\cal W}_{bp;x,y} \rangle = 0$ if $a \neq b$, which provides an alternative proof of Eq.\ (\ref{eq:WW}) of the main text.

To arrive at Eq.\ (\ref{eq:Wgeneral}) for the expectation value $\langle |{\cal W}_{ap;x,y}|^2 \rangle$ we combine all 24 contributions shown in Fig.\ \ref{fig:diagrams}, evaluate the traces of the products of generalized Pauli matrices as described above, and use the relations between the coefficients $B_j$ given in Eqs.\ (\ref{eq:Bconstraint1})--(\ref{eq:Bconstraint3}) to express all coefficients in terms of $B_{5}$, $B_{3}$, and $B_{6}$. The resulting expression is of the form (\ref{eq:Wgeneral}) with
\begin{align}
  \label{eq:Aresult}
  A_{1} =&\, \frac{2 ({\cal R}_{2;2} + {\cal R}_{1,1;1,1} - 4 {\cal R}_{1,1})}{q^4 (q^4 - 9)} 
  \nonumber \\ &\, \mbox{}
- \frac{2 (q^4-3)\mbox{Re}\, {\cal R}_{1,1;2}}{q^6(q^4-9)} - i \frac{2 \mbox{Im}\, {\cal R}_{1,1;2}}{q^2(q^4-4)}, \\
  A_{2} =&\, \frac{(q^4 + 6)({\cal R}_{2;2} + {\cal R}_{1,1;1,1} - 4 {\cal R}_{1,1})}{q^4 (q^4 - 9)(q^4-4)} \nonumber \\ &\, \mbox{}
  + \frac{{\cal R}_{2;2} - 2}{q^4-4} - \frac{2 \mbox{Re}\, {\cal R}_{1,1;2}}{q^2(q^4-9)},
  \\
  A_{3} =&\, 
  \frac{(q^8 - 8 q^4 + 6)({\cal R}_{2;2} + {\cal R}_{1,1;1,1} - 4 {\cal R}_{1,1})}{q^4 (q^4 - 9)(q^4-4)} \nonumber \\ &\, \mbox{}
  - \frac{{\cal R}_{2;2}-2}{q^4-4} - \frac{2 \mbox{Re}\,{\cal R}_{1,1;2}}{q^2 (q^4-9)}.
  \label{eq:A3result}
\end{align}
The coefficient $A_1$ is complex, whereas $A_2$ and $A_3$ are real.

\section{Pauli-string weights $\rho_p(t)$ of binary form}
\label{app:3}

We show that if $\rho_a(t-1)$ is of binary form, then $\rho_{a}(t)$ is binary, too, for $t$ even. The proof for $t$ odd is analogous.

Since the evolution described by Eq.\ (\ref{eq:gammasqavg}) evolves qudit pairs separately, it is sufficient to consider a single qudit pair at positions $x$ and $x+1$, $x$ even. We split the summation over $a_x$ and $a_{x+1}$ into four parts,
\begin{align}
  \label{eq:gammatsum}
  \lefteqn{
  \rho_{p_x,p_{x+1}}(t) =
  \rho_{0,0}(t-1)
  \langle |W_{0,0,p_x,p_{x+1}}|^2 \rangle
} ~~~~
  \\ &\, \mbox{}
  + \sum_{a_x \neq 0}
  \rho_{a_x,0}(t-1)
  \langle |W_{a_x,0,p_x,p_{x+1}}|^2 \rangle
  \nonumber \\ &\, \mbox{}
  + \sum_{a_{x+1} \neq 0}
  \rho_{0,a_{x+1}}(t-1)
  \langle |W_{0,a_{x+1},p_x,p_{x+1}}|^2 \rangle
  \nonumber \\ &\, \mbox{}
  + \sum_{a_x \neq 0} \sum_{a_{x+1} \neq 0}
  \rho_{a_{x},a_{x+1}}(t-1)
  \langle |W_{a_{x},a_{x+1},p_x,p_{x+1}}|^2 \rangle. \nonumber
\end{align}
Since $\rho_{a}(t-1)$ is binary,
it is a constant in each of the four lines in Eq.\ (\ref{eq:gammatsum}). The transition probability $\langle |W_{0,0,p_x,p_{x+1}}|^2 \rangle$, as well as the summations $\sum_{a_x \neq 0} \langle |W_{a_x,0,p_x,p_{x+1}}|^2 \rangle$, $\sum_{a_{x+1} \neq 0} \langle |W_{0,a_{x+1},p_x,p_{x+1}}|^2 \rangle$, and $\sum_{a_x \neq 0} \sum_{a_{x+1} \neq 0}  \langle |W_{a_{x},a_{x+1},p_x,p_{x+1}}|^2 \rangle$ depend on $p_x$ and $p_{x+1}$ through the Kronecker deltas $\delta_{p_x,0}$ and $\delta_{p_{x+1},0}$ only. This follows from Eq.\ (\ref{eq:Wgeneral}), using the properties (\ref{eq:phispecial}) and (\ref{eq:phiidentities}) of the function $\varphi(a,p)$. Since 
$\rho_{p_{x},p_{x+1}}(t)$
depends on $p_x$ and $p_{x+1}$ through the Kronecker deltas $\delta_{p_x,0}$ and $\delta_{p_{x+1},0}$ only, it has the binary form.

\section{Drift-diffusion equation}
\label{app:5}

Inspired by Refs.\ \cite{gcrw, robin1984two, Eric1981one}, we write the two-component spinor $R_{\rm R}^{(n)}(\Delta x,t)$, which represents the $n$-point densities, in the basis of right-eigenvectors of $D^{(n)} + D'^{(n)}$,
\begin{equation}
  R_{\rm R}^{(n)}(\Delta x,t) = \sum_{i=1}^{2^{n+1}} R_{{\rm R}i}^{(n)}(\Delta x,t) V_i^{(n)}.
\end{equation}
Assuming that $R_{\rm R}^{(n)}(\Delta x,t)$ is a sufficiently smooth function of $\Delta x$ and $t$, we expand to first order in changes in $\Delta x$ and $t$, and find that the expansion coefficients satisfy the equations
\begin{align}
  (1 - d_i^{(n)}) R_{{\rm R}i}^{(n)}(\Delta x,t) =&\,
  - \partial_t R_{{\rm R}i}^{(n)}(\Delta x,t) 
   \\ &\, \mbox{}\nonumber
  + 2 \sum_{j} d'^{(n)}_{ij} \partial_{\Delta x} R_{{\rm R}j}(\Delta x,t),
  \label{eq:Rj}
\end{align}
where $d_{ij}'^{(n)}$ was defined in Eq.\ (\ref{eq:dijdef}).
Anticipating that the expansion coefficients $R_{{\rm R}i}^{(n)}(\Delta x,t)$ with $i > 1$ are parametrically smaller than $R_{{\rm R}1}^{(n)}(\Delta x,t)$ in the long-time limit, which will be shown below, we focus on the evolution equation for $R_{{\rm R}1}^{(n)}(\Delta x,t)$. Substituting $d_1^{(n)} = 1$ gives
\begin{equation}
  0 = (- \partial_t  + 2 d_{11}'^{(n)} \partial_{\Delta x})
  R_{{\rm R}1}^{(n)}(\Delta x,t).
  \label{eq:R10}
\end{equation}
This equation determines the Butterfly velocity $v_{\rm B}^{(n)}$, see Eq.\ (\ref{eq:vBn}).


To capture the long-time diffusive spreading of the propagating front, we switch to coordinates co-moving with the propagating front. Formally, this is achieved by the definition
\begin{equation}
  Q_{{\rm R}j}^{(n)}(\Delta y,t) = R_{{\rm R}j}^{(n)}(\Delta y +2 d_{11}^{(n)} t,t),
\end{equation}
We first consider the expansion coefficients $Q_{{\rm R}j}^{(n)}(\Delta y,t)$ with $j > 1$. Their evolution equation reads
\begin{align}
  (1 - d_{\rm j}^{(n)}) Q_{{\rm R}j}^{(n)}(\Delta y,t) =&\,
  2 d_{j1}'^{(n)} \partial_{\Delta y} Q_{{\rm R}1}^{(n)}(\Delta y,t),
  \label{eq:Qj}
\end{align}
where we dropped all terms involving derivatives of $Q_{{\rm R}j}(\Delta y,t)$ for $j > 1$. Solving Eq.\ (\ref{eq:Qj}) gives
\begin{equation}
  Q_{{\rm R}j}^{(n)}(\Delta y,t) =
  \frac{2 d_{j1}'^{(n)}}{1 - d_{\rm j}^{(n)}} \partial_{\Delta y} Q_{{\rm R}1}^{(n)}(\Delta y,t).
  \label{eq:Qjsol}
\end{equation}
To find an evolution equation for $Q_{{\rm R}1}^{(n)}(\Delta y,t)$ that accounts for the diffusive spreading of the front, we now keep the terms involving first-order spatial derivatives of $Q_{{\rm R}j}^{(n)}(\Delta y,t)$ with $j > 1$ as well as terms with a second-order spatial derivative of $Q_{{\rm R}1}^{(n)}(\Delta y,t)$. As a result, one finds that $Q_{{\rm R}1}^{(n)}(\Delta y,t)$ obeys the diffusion equation,
\begin{align}
  \partial_t Q_{{\rm R}1}^{(n)}(\Delta y,t) =&\, 
 \frac{ {\cal D}^{(n)}}{2}
  \partial_{\Delta y}^2 Q_{{\rm R}1}^{(n)}(\Delta y,t),
  \label{eq:diffusion}
\end{align}
with the diffusion constant ${\cal D}^{(n)}$ given by Eq.\ (\ref{eq:DDn}). (There is no term with a first-order spatial derivative since we use the coordinate $\Delta y$ co-moving with the front.) Replacing the continuous variable $\Delta y$ by $\Delta x = \Delta y + 2 d_{11}'^{(n)} t$ gives the drift-diffusion equation (\ref{eq:driftdiffusion}) of the main text.

It remains to verify that $R_{{\rm R}}^{(n)}(\Delta x,t)$ is dominated by the expansion coefficient $R_{{\rm R}1}^{(n)}(\Delta x,t)$ in the long-time limit, which was an essential assumption in our derivation of Eq.\ (\ref{eq:diffusion}). Hereto, we note that the solution of the diffusion equation (\ref{eq:diffusion}) is $Q_{{\rm R}1}^{(n)}(\Delta y,t) \propto \mbox{erfc}(\Delta y/2 \sqrt{{\cal D}^{(n)} t})$. Its derivative to $\partial_{\Delta y} Q_{{\rm R}1}^{(n)}(\Delta y,t)$ is proportional to $1/\sqrt{t}$. Hence, by Eq.\ (\ref{eq:Qjsol}), $Q_{{\rm R}j}^{(n)}(\Delta y,t)$ is indeed parametrically smaller than $Q_{{\rm R}1}^{(n)}$ in the long-time limit.

The above construction of the drift-diffusion equation for $R_{{\rm R}1}^{(n)}(\Delta x,t)$ makes use of the assumption that a complete basis of left- and right-eigenvectors $\tilde V_j^{(n)}$ and $V_j^{(n)}$ of $D^{(n)} + D'^{(n)}$ exists. With the exception of the requirement that left- and right-eigenvectors $\tilde V_1^{(n)}$ and $V_1^{(n)}$ exist, which is guaranteed, because $\tilde V_1^{(n)} = (1,1,\ldots,1)^{\rm T}$ can be explicitly constructed, this assumption is not necessary for the mapping to the drift-diffusion equation (\ref{eq:driftdiffusion}). The extension of the above construction to the case that a complete basis of left- and right-eigenvectors of $D^{(n)} + D'^{(n)}$ does not exist is straightforward and one finds the diffusion constant
\begin{align}
  {\cal D}^{(n)} =&\, 4 d_{11}'^{(n)}[1 - d_{11}'^{(n)}]
  \nonumber \\ &\, \mbox{}
  + 8\tilde V_1^{(n){\rm T}}
  D'^{(n)} 
  G^{(n)}
  D'^{(n)} V_1^{(n)},
\end{align}
with
\begin{align}
  G^{(n)} =&\, 
  \sum_{l=0}^{\infty}
  (D^{(n)} + D'^{(n)})^l (\openone - 
  V_1^{(n)} \tilde V_1^{(n){\rm T}}) \nonumber \\ =&\,
\left[ \openone - (D^{(n)} + D'^{(n)})(\openone -
  V_1^{(n)} \tilde V_1^{(n){\rm T}})\right]^{-1} 
  \nonumber \\ &\, \mbox{} - 
  V_1^{(n)} \tilde V_1^{(n){\rm T}}.
\end{align}

\section{Brownian-motion ensemble}
\label{app:Brownian}
\textcolor{black}{
The Poisson kernel offers a continuous interpolation between a random circuit with Haar-distributed two-qudit evolution operators and the trivial circuit. Another random matrix ensemble that interpolates between these two limits is Dyson's Brownian motion ensemble \cite{Dyson1962Brownian,Dyson1972Brownian}. It obtains the two-qudit operators ${\cal U}_{x,x+1}$ from the continuous time evolution with a random Hamiltonian with Gaussian white-noise distribution. Specifically,
\begin{equation}
  {\cal U} = T_t e^{-i \int_0^1 dt {\cal H}(t)},
\end{equation}
where $T_t$ indicates the time-ordering and ${\cal H}(t')$ is a $q^2 \times q^2$ hermitian matrix with Gaussian distribution with zero mean and variance
\begin{equation}
  \langle H_{ij}(t) H_{kl}(t') \rangle =
  \frac{\lambda}{q^2}
  \delta_{il} \delta_{jk} \delta(t-t'),
\end{equation}
where $\lambda$ is a parameter that describes the interpolation between the trivial and Haar-random two-qudit gate operators. This ensemble satisfies the unitary invariance property (\ref{eq:unitary_invariance}). The distribution of the eigenphases $e^{i \theta_j}$ of ${\cal U}$, $j=1,2,\ldots,q^2$, is \cite{Pandey1991}
\begin{align}
  P(\theta_1,\ldots,\theta_{q^2}) =&\,
  \frac{2^{\frac{1}{2}q^2(q^2-1)}}{(q^2)!} 
  e^{\frac{\lambda}{24} (q^4-1)}
  \prod_{j < k} \sin \frac{\theta_j - \theta_k}{2}
\nonumber \\ &\, \mbox{} \times  \det\left[ \frac{f^{(j)}(\theta_i)}{j!} \right]_{i,j=1,\ldots,q^2},
\end{align}
where
\begin{align}
  f(\theta) =&\, \frac{1}{2 \pi} \sum_{k=-\infty}^{\infty}
  e^{-k^2 \lambda/2 q^2 + i k \theta}
  \nonumber \\
  =&\,
  \sqrt{\frac{q^2}{2 \pi \lambda}}
  \sum_{k=-\infty}^{\infty}
  e^{-q^2 (\theta - 2 \pi k)^2/2 \lambda}
\end{align}
and $f^{(j)}$ denotes the $j$th derivative.}

\textcolor{black}{The moments ${\cal R}_{1,1}$, ${\cal R}_{1,1;1,1}$, ${\cal R}_{2;2}$, and ${\cal R}_{1,1;2}$ are \cite{tang2024brownian} 
\begin{align}
 \mathcal{R}_{1;1} =&\, q^4 e^{-\lambda} + 1 - e^{-\lambda}, \\
    \mathcal{R}_{2;2} =&\, 2 + \frac{1}{2} e^{-2 \lambda}
  \left( q^4 (q^4-3) \cosh \frac{2 \lambda}{q^2}
  \right. \nonumber \\ &\, \left. \ \ \ \ \mbox{} -
  2 q^6 \sinh \frac{2 \lambda}{q^2} - q^8 + 5 q^4 - 4 \right),
  \\
    \mathcal{R}_{1,1;2} =&\,
  e^{-2 \lambda} q^4 \left[ q^2 \cosh\frac{2 \lambda}{q^2}
  + \sinh \frac{2 \lambda}{q^2} \right], \\
    \mathcal{R}_{1,1;1,1} =&\, 
  {\cal R}_{2;2} +  4 (q^4 -1) e^{-\lambda}+(q^8-5q^4+4)e^{-2\lambda}.
\end{align}
From Eqs.\ (\ref{eq:T}), (\ref{eq:vB0result}), (\ref{eq:DD0result}), and (\ref{eq:Aresult})--(\ref{eq:A3result}), we then find that the lowest-order approximations for the butterfly velocity and the diffusion constant are 
\begin{align}
  v_{\rm B}^{(0)} =&\,\frac{q^2-1}{q^2+1}\frac{c-d}{c+d} \\
  {\cal D}^{(0)} =&\,\frac{4}{(q^2+1)^2}\frac{(c-d)(c+q^2d)(q^2c+d)}{c(c+d)^2},
\end{align}
where
\begin{align}
    c=&\, e^{2\lambda}(q^6-9q^2)\\
    d=&\, (q^6-9q^2)\cosh (\frac{2\lambda}{q^2})-2(q^4+3)\sinh(\frac{2\lambda}{q^2}).
\end{align}
In the limit $q \to \infty$ these estimates agree with what we find for the Poisson kernel distribution, provided we identify $|\alpha|^2 = e^{-\lambda}$.}

\textcolor{black}{The Brownian motion ensemble can also be used to construct a continuous random process. 
To find the drift velocity and the diffusion constant within our approach, we rescale $t' = \lambda t$ and take the limit $\lambda \to 0$. Using $t'$ as the time variable for the random continuous process, the lowest-order approximations for the drift velocity and diffusion constant become
\begin{align}
  v_{\rm B}'^{(0)} =&\, \frac{(q^2-1)^2(q^4-6)}{q^4(q^4-9)}, \\
  {\cal D}'^{(0)} =&\, \frac{2(q^4-1)(q^4-6)}{q^4(q^4-9)}.
\end{align} 
For large $q$, these expressions agree with the results obtained for the Poisson kernel, Eqs.\ (\ref{eq:vBcontinuous}) and (\ref{eq:Dcontinuous}).
}
\vskip 0.5 cm
\section{Poisson kernel}
\label{app: 8}

To describe operator spreading in a random unitary circuit with Poisson-kernel-distributed two-qudit gate operators, we need to calculate the moment functions ${\cal R}_{1;1}$, ${\cal R}_{1,1;1,1}$, ${\cal R}_{2;2}$, and ${\cal R}_{1,1;2}$ for the Poisson kernel distribution. The moment functions are defined in Eq.\ (\ref{eq:R112}). 

A unitary matrix ${\cal U}$ distributed according to the Poisson kernel distribution (\ref{eq:Poisson}) may be parameterized as \cite{brouwer_generalized_1995,mello1985a},
\begin{equation}
  \mathcal{U}= (\alpha \openone - {\cal U}_0)(\openone - \alpha^* {\cal U}_0)^{-1},
 \label{eq:UU0}
\end{equation}
where $\mathcal{U}_0$ is drawn from a Haar distribution. Using the known distribution of the eigenphases $e^{i \phi_j}$, $j=1,\ldots,q^2$ of a Haar-distributed $q^2 \times q^2$ unitary matrix \cite{mehta2004random,forrester2010},
\begin{equation}
  P_0(\phi_1,\ldots,\phi_{q^2}) =
  \frac{1}{q^2! (2 \pi)^{q^2}}
  \prod_{i < j} |e^{i \phi_i} - e^{i \phi_j}|^2,
\end{equation}
and using Eq.\ (\ref{eq:UU0}), we find
\begin{align}
  {\cal R}_{1;1} =&\,
  1 + |\alpha|^2 q^4 - |\alpha|^{2 q^2}, \label{eq: R11} \\
  \mathcal{R}_{2;2} =&\,
  2 + |\alpha|^4 q^4 - |\alpha|^{2 q^2 - 2} (1 - |\alpha|^2)^2 q^4
  \nonumber \\ &\, \mbox{} 
  - 2 |\alpha|^{2 q^2}, \label{eq: R22} \\
  \mathcal{R}_{1,1;1,1} =&\,
  2 + 4 |\alpha|^2 q^4 + |\alpha|^4 q^8
  \nonumber \\ &\, \mbox{} 
   - |\alpha|^{2 q^2-2} q^4 (1 + |\alpha|^2)^2
  - 2 |\alpha|^{2 q^2}, \label{eq: R1111} \\
  \mathcal{R}_{1,1;2} =&\, |\alpha|^4 q^6 + |\alpha|^{2 q^2-2} q^4 (1 - |\alpha|^4). \label{R112}
\end{align}
For the coefficients $A_1$, $A_2$, and $A_3$ this gives Eqs.\ (\ref{eq:APoisson1})--(\ref{eq:APoisson3}) from the main text, when using Eqs.\ (\ref{eq:Aresult})--(\ref{eq:A3result}).

We now discuss the derivation of these results. Expanding Eq.\ (\ref{eq:UU0}), we write the Poisson-kernel-distributed unitary matrix ${\cal U}$ as
\begin{align}
    \mathcal{U}=&\, \alpha \openone-(1-|\alpha|^2)\sum_{l=0}^{\infty}\alpha^{*l}\mathcal{U}_0^{l+1}, \label{eq: Uexpand}
\end{align}
where ${\cal U}_0$ is Haar-distributed.
Using Eq. (\ref{eq: Uexpand}), one finds
\begin{widetext}
\begin{align}
    \mathcal{R}_{1,1}=&\, \left\langle \left(\alpha^* q^2-(1-|\alpha|^2)\sum_{l=0}^{\infty}\alpha^l\tr \mathcal{U}_0^{l+1}\right)\left(\alpha q^2-(1-|\alpha|^{2})\sum_{l=0}^{\infty}\alpha^{*l}\tr (\mathcal{U}_0^{\dagger})^{l+1}\right) \right\rangle \nonumber \\
    =&\, |\alpha|^2 q^4 + (1-|\alpha|^2)^2\sum_{l=0}^{\infty}\sum_{m=0}^{\infty}\alpha^{l}\alpha^{*m}\left<\tr \mathcal{U}_0^{l+1}\tr (\mathcal{U}_0^{\dagger})^{m+1}\right>,
\end{align}
where we used that $\tr \openone = q^2$ for a matrix of size $q^2 \times q^2$.
From Refs.\ \cite{diaconis2001linear} and \cite{johansson1997random}, we know that that the ensemble average
\begin{equation}
   \langle \tr U_0^{j} \tr U_0^{\dagger k} \rangle =
   \delta_{j,k} \mathrm{min} \{j, q^2\}.\label{eq: twoU}
\end{equation}
Therefore, we find
\begin{align}
\mathcal{R}_{1;1}=& |\alpha|^2 q^4+ (1-|\alpha|^2)^2\sum_{l=0}^{q^2-1}|\alpha|^{2l}(l+1)+q^2(1-|\alpha|^2)^2\sum_{l=q^2}^{\infty}|\alpha|^{2l}\nonumber\\
=&1+|\alpha|^2q^4-|\alpha|^{2q^2},
\end{align}
which reproduces Eq.\ (\ref{eq: R11}).
Following the same argument, one finds that
\begin{align}
 \mathcal{R}_{2;2} =&\, |\alpha|^4 q^4+4|\alpha|^2(1-|\alpha|^2)^2\sum_{l,m=0}^{\infty} \alpha^{l}\alpha^{*m}\left<\tr \mathcal{U}_0^{l+1}\tr (\mathcal{U}_0^{\dagger})^{m+1}\right> \nonumber \\ &\, \mbox{}
+(1-|\alpha|^2)^4\sum_{m,l,k,n=0}^{\infty}\alpha^{k+l}(\alpha^*)^{m+n}\nonumber\left<\tr \mathcal{U}_0^{k+l+2}\tr (\mathcal{U}_0^{\dagger})^{m+n+2}\right>
 \nonumber \\ &\, \mbox{} -2(1-|\alpha|^2)^3\sum_{l,m,n=0}^{\infty}\alpha^l(\alpha^*)^{m+n+1}\left<\tr \mathcal{U}_0^{l+1}\tr (\mathcal{U}_0^{\dagger})^{m+n+2}\right>
 \nonumber \\ &\, \mbox{}
-2(1-|\alpha|^2)^3\sum_{l,m,n=0}^{\infty}\alpha^{*l}(\alpha)^{m+n+1}\nonumber \left<\tr \mathcal{U}_0^{m+n+2}\tr (\mathcal{U}_0^{\dagger})^{l+1}\right>,
\end{align}
which gives Eq.\ (\ref{eq: R22}) upon substitution of Eq.\ (\ref{eq: twoU}).

Likewise, for the two remaining moments ${\cal R}_{1,1;1,1}$ and ${\cal R}_{1,1;2}$ we find
\begin{align}
\mathcal{R}_{1,1;1,1}=&\, |\alpha|^4 q^8+4|\alpha|^2(1-|\alpha|^2)^2 q^4 \nonumber \sum_{l,m=0}^{\infty} \alpha^{l}\alpha^{*m}\left<\tr \mathcal{U}_0^{l+1}\tr (\mathcal{U}_0^{\dagger})^{m+1}\right> \nonumber \\ &\, \mbox{}
+(1-|\alpha|^2)^4\sum_{m,l,k,n=0}^{\infty}\alpha^{k+l}(\alpha^*)^{m+n}\left<\tr \mathcal{U}_0^{k+1}\tr \mathcal{U}_0^{l+1}\tr (\mathcal{U}_0^{\dagger})^{m+1}\tr (\mathcal{U}_0^{\dagger})^{n+1}\right> \nonumber \\ &\, \mbox{}
-2(1-|\alpha|^2)^3 q^2\sum_{l,m,n=0}^{\infty}\alpha^l(\alpha^*)^{m+n+1}\left<\tr \mathcal{U}_0^{l+1}\tr (\mathcal{U}_0^{\dagger})^{m+1}\tr (\mathcal{U}_0^{\dagger})^{n+1}\right>\nonumber \\ &\, \mbox{}
-2(1-|\alpha|^2)^3 q^2\sum_{l,m,n=0}^{\infty}\alpha^{*l} \alpha^{m+n+1}\left<\tr \mathcal{U}_0^{m+1}\tr \mathcal{U}_0^{n+1}\tr (\mathcal{U}_0^{\dagger})^{l+1}\right>,
\end{align}
and
\begin{align}
\mathcal{R}_{1,1;2}=&\, |\alpha|^4 q^6+4|\alpha|^2(1-|\alpha|^2)^2 q^2\sum_{l,m=0}^{\infty} \alpha^{l}\alpha^{*m}\left<\tr \mathcal{U}_0^{l+1}\tr (\mathcal{U}_0^{\dagger})^{m+1}\right> \nonumber \\ &\, \mbox{}
+(1-|\alpha|^2)^4\sum_{m,l,k,n=0}^{\infty}\alpha^{k+l}(\alpha^*)^{m+n}\left<\tr \mathcal{U}_0^{k+1}\tr \mathcal{U}_0^{l+1}\tr (\mathcal{U}_0^{\dagger})^{m+n+2}\right> \nonumber \\ &\, \mbox{}
-2(1-|\alpha|^2)^3 q^2\sum_{l,m,n=0}^{\infty}\alpha^l(\alpha^*)^{m+n+1}\left<\tr \mathcal{U}_0^{l+1}\tr (\mathcal{U}_0^{\dagger})^{m+n+2}\right>\nonumber \\ &\, \mbox{}
-2(1-|\alpha|^2)^3\sum_{l,m,n=0}^{\infty}\alpha^{*l}(\alpha)^{m+n+1} \left<\tr \mathcal{U}_0^{m+1}\tr \mathcal{U}_0^{n+1}\tr (\mathcal{U}_0^{\dagger})^{l+1}\right>.
\end{align}
Ensemble averages containing a product of two traces can again be calculated with the help of Eq.\ (\ref{eq: twoU}). 
Since $\langle \tr U_0^{j_1}\tr U_0^{j_2}\tr U_0^{\dagger k_1}\tr U_0^{\dagger k_2} \rangle \propto \delta_{j_1+j_2,k_1+k_2}$, one can simplify the remaining averages in the expressions for $\mathcal{R}_{1,1;1,1}$ and $\mathcal{R}_{1,1;2}$ as
\begin{align}
\mathcal{R}_{1,1;1,1}=&\, |\alpha|^4 q^8 +4|\alpha|^2q^4(1-|\alpha|^{2q^2}) 
  +(1-|\alpha|^2)^3\sum_{K=2}^{\infty}|\alpha|^{2(K-2)}
  [ (1 - |\alpha|^2) {\cal Q}(K) 
  - 4 |\alpha|^2 q^2 \mbox{Re}\, {\cal Q}'(K) ]
\end{align}
and 
\begin{align}
  \mathcal{R}_{1,1;2}=&\, |\alpha|^4 q^6 
  + 2 q^4 |\alpha|^{2 q^2}(1 - |\alpha|^2))
  + (1-|\alpha|^2)^3\sum_{K=2}^{\infty}|\alpha|^{2(K-2)} 
  (K-1 - |\alpha|^2(K+1))
  {\cal Q}'(K),
\end{align}
\end{widetext}
where we abbreviated
\begin{align} \label{eq: fourpoint}
  {\cal Q}(K)  =&\, \sum_{l,m=1}^{K-1}\left<\tr \mathcal{U}_0^{K-l} \tr \mathcal{U}_0^{l}\tr (\mathcal{U}_0^{\dagger})^{m}\tr (\mathcal{U}_0^{\dagger})^{K-m}\right>\nonumber ,
\end{align}
\begin{align}
  {\cal Q}(K)' =&\, \sum_{m=1}^{K-1}
  \left<\tr \mathcal{U}_0^{K-m}\tr \mathcal{U}_0^{m}\tr (\mathcal{U}_0^{\dagger})^{K}\right>.
\end{align}
An expression for the average of a product of traces of powers of ${\cal U}_0$ is given in Refs.\ \cite{diaconis2001linear} and \cite{johansson1997random}, 
\begin{align}
  \langle \tr U_0^{j_1}\tr U_0^{j_2}\tr U_0^{\dagger k_1}\tr U_0^{\dagger k_2} \rangle =&\, \sum_{ \stackrel{\lambda \vdash K}{L(\lambda)\le q^2}} \chi_{\lambda} (\iota)\chi_{\lambda}(\kappa),
 \label{eq: four_char}
\end{align}
where the summation is over partitions $\lambda$ of $K=j_1+j_2=k_1+k_2$, $L(\lambda)$ is the length of $\lambda$, and $\chi_{\lambda}(\mu)$ is the irreducible characters of the symmetric group $S_K$. (Recall that the size of the unitary matrices we consider is $q^2$. A partition $\lambda$ of $K$ consists of the positive integers $\lambda_1 + \lambda_2 + \ldots + \lambda_L = K$, with $L \equiv L(\lambda)$ and $\lambda_1 \ge \lambda_2 \ge \ldots \ge \lambda_L$.) Note that $j_1 + j_2 = K$ and $k_1 + k_2 = K$ are partitions of $K$, too. We use $\iota$ and $\kappa$ to denote these partitions, whereby the ordering of the two terms in the partition is chosen such that the first term is $\ge$ the second one. (Explicitly, $\iota = (j_1,j_2)$ if $j_1 \ge j_2$ and $(j_2,j_1)$ otherwise, and $\kappa = (k_1,k_2)$ if $k_1 \ge k_2$ and $(k_2,k_1)$ otherwise.)

The irreducible characters of the symmetric group can be obtained from Frobenius formula \cite{fulton_representation_2004}, which states that $\chi_{\lambda}(\iota)$ is the coefficient of the monomial $x_1^{l_1} x_2^{l_2} \ldots x_{L}^{l_{L}}$ in 
$$
  \prod_{i<j}(x_i-x_j) P_{j_{1}}(x)P_{j_2}(x),
$$%
where $P_{j}(x)=\sum_{i=1}^{L} x_{i}^{j}$ and $l_i=\lambda_i+L-i$. 
The product $\prod_{i<j}(x_i-x_j) $ is the Vandermonde determinant
\begin{equation}
    \prod_{i<j}(x_i-x_j)=\sum_{\sigma\in S_L}(\mathrm{sgn}\ \sigma)x_L^{\sigma(1)-1}\cdots x_1^{\sigma(L)-1}.
\end{equation}
Because $l_1\ge L$, the only non-vanishing terms are
\begin{align*}
   &\mathrm{sgn}(\sigma) x_L^{\sigma(1)-1}\cdots x_1^{j_1+j_2+\sigma(L)-1} \\
   & \mathrm{sgn}(\sigma)x_L^{\sigma(1)-1}\cdots x_i^{j_2+\sigma(L+1-i)-1}\cdots x_1^{j_1+\sigma(L)-1}\\
   &\mathrm{sgn}(\sigma) x_L^{\sigma(1)-1}\cdots x_i^{j_1+\sigma(L+1-i)-1}\cdots x_1^{j_2+\sigma(L)-1}
\end{align*}

The calculation of $\chi_{\lambda}(\iota)$ depends on whether $\lambda$ is a hook partition, {\em i.e.}, a partition of the form $\lambda_1 + 1 + \ldots + 1$, in which only one term is $> 1$. If $\lambda$ is hook, then 
\begin{align}
    \chi_{\lambda}(\iota) &=(-1)^{L}, \ & j_1< L\le K; \nonumber\\
     \chi_{\lambda}(\iota) &=0, \ \quad & j_2<L\le j_1;\nonumber\\
      \chi_{\lambda}(\iota) &=(-1)^{L-1}, \ & 1\le L\le j_2.
\end{align}

If $\lambda$ is not a hook partition, then for fixed $\iota$ and $\kappa$ there is only one permutation $\sigma$ that gives a non-vanishing value, which is $(2-\mathrm{sgn}(|j_1-j_2|))\mathrm{sng}(\sigma)$. One can further prove that for each partition $\lambda$ there are only two partitions $\iota$ that generate nonzero $\chi_{\lambda}(\iota)$. To see this, we suppose that $\iota=(j_1,j_2)$ and $\iota'=(j'_1,j'_2)$ are two solutions that generate nonzero $\chi_{\lambda}(\iota)$ and $\chi_{\lambda}(\iota')$ for the same partition $\lambda$. The non-vanishing terms are
\begin{align*}
  & \mathrm{sgn}(\sigma) x_L^{\sigma(1)-1}\cdots x_i^{j_2+\sigma(L+1-i)-1}\cdots x_1^{j_1+\sigma(L)-1}\\
   &\mathrm{sgn}(\sigma')x_L^{\sigma'(1)-1}\cdots x_{i'}^{j'_2+\sigma'(L+1-i')-1}\cdots x_1^{j'_1+\sigma'(L)-1}
\end{align*}
Because each part of $\lambda$ is positive, one must have either $\sigma(L)=1$ or $\sigma(L+1-i)=1.$ If $\sigma(L)=1$, then we must have $\sigma'(L+1-i')=1$. We can rule out the possibility $i\ne i'$ because if so we must have $\sigma'(L)=j'_2+1$. In this case, $\lambda$ is a hook, which contradicts our presumption. Moreover, it also means there is no third partition $\iota'': j_1''+j_2''=K$ to generate nonzero $\chi_\lambda(j_1'',j_2'')$. We then can further conclude that $\sigma(L+1-i)=\sigma'(L)=j_1-j'_1+1.$  For other terms, $\sigma$  and $\sigma'$ are the same. Therefore, the sign of $\sigma$ and $\sigma'$ are opposite. Because $j_1 \ge j_2$ and $j_2 \ge j_1$ correspond to the same partition, we have 
\begin{equation}
    \sum_{j_1=1}^{K}\chi_{\lambda}(j_1,j_2)=0
\end{equation}
if $\lambda$ is not a hook partition.
Furthermore, for a hook partition,
\begin{equation}
    \sum_{j_1=1}^{K-1}\chi_{\lambda}(\iota) =(-1)^{L(\lambda)}(2L(\lambda)-K-1)
\end{equation}
Using Eq.\ (\ref{eq: four_char}) to calculate the average (\ref{eq: fourpoint}), we obtain  
\begin{align}
  {\cal Q}(K) =&\, \sum_{j_1=1}^{K-1}\sum_{k_1=1}^{K-1} \langle \tr U_0^{j_1}\tr U_0^{j_2}\tr U_0^{\dagger k_1}\tr U_0^{\dagger k_2} \rangle \nonumber \\
& =\sum_{\stackrel{\lambda \vdash K}{L(\lambda)\le q^2}}\sum_{j_1=1}^{K-1} \chi_{\lambda}(\iota) \sum_{k_1 = 1}^{K-1} \chi_{\lambda}(\kappa)   \nonumber\\
&=\sum_{L(\lambda)=1}^{\mathrm{min}\{q^2,K\}}(2L(\lambda)-K-1)^2\nonumber\\
&= \begin{cases} \frac{1}{3} 
    q^2(4q^4-6K q^2+3K^2-1) & \mbox{if $q^2\le K$}, \\
    \frac{1}{3} K(K^2-1),
    \ & \mbox{if $q^2>K$}.
\end{cases}
\end{align}
Similarly, we find
\begin{align}
  {\cal Q}'(K) =&\,
    \sum_{k_1=1}^{K-1} \langle \tr U_0^{K}\tr U_0^{\dagger k_1}\tr U_0^{\dagger k_2} \rangle \nonumber \\
    =&\,\sum_{\stackrel{\lambda \vdash K}{L(\lambda) \le q^2}} \chi_{\lambda} ((K))\sum_{k_1 = 1}^{K-1} \chi_{\lambda}(\kappa)  \nonumber \\
    =&\, \sum_{L(\lambda)=1}^{\mathrm{min}\{q^2,K\}}-(2L(\lambda)-K-1)
  \nonumber \\ =&\,
  \begin{cases}
        q^2(K-q^2),\ & \mbox{if $q^2\le K$}, \\
        0 & \mbox{if $q^2 > K$},
    \end{cases} 
\end{align}
where $(K)$ denotes the length-one partition $K = K$. Upon substituting these results, we obtain Eq.\ (\ref{eq: R1111}) and Eq.\ (\ref{R112}).

\bibliography{butterfly}
\end{document}